\documentclass[11pt, letterpaper]{article}

\usepackage[left=1in, right=1in, top=1in, bottom=1in]{geometry}
\usepackage[utf8]{inputenc}
\usepackage[T1]{fontenc}
\usepackage{bm}
\usepackage{type1cm}
\usepackage{lettrine}
\usepackage{amsmath,amssymb,amsthm}
\usepackage{moreverb}
\usepackage{mathtools}
\usepackage{amsmath}
\usepackage{svg}
\usepackage{amssymb}
\usepackage{algorithmic}
\usepackage{graphics}
\usepackage{graphicx}
\usepackage{subfigure}
\usepackage{caption}
\usepackage{extarrows}
\usepackage{color}
\usepackage{framed}
\usepackage{wrapfig}
\usepackage{algorithm}
\usepackage{algorithmic}
\usepackage{bm}
\usepackage{mathrsfs}
\usepackage{mathabx}
\usepackage{multirow}
\usepackage{longtable}
\usepackage{hyperref}
\usepackage{paralist}
\usepackage{indentfirst}
\usepackage{relsize}
\usepackage{extarrows}
\usepackage{upgreek}
\usepackage{bm}
\usepackage{mwe}
\usepackage[dvipsnames,table]{xcolor}
\usepackage{booktabs}
\usepackage{authblk}
\usepackage{lettrine}
\usepackage{type1cm}
\usepackage{threeparttable}
\usepackage[sort&compress,numbers]{natbib}
\usepackage[figurename=Figure]{caption}
\usepackage[normalem]{ulem}
\usepackage{adjustbox}
\usepackage{ulem}
\usepackage{xcolor}
\usepackage{amsmath,amsfonts,bm}

\def\eqref#1{equation~\ref{#1}}
\def\1{\bm{1}}

\DeclareMathAlphabet{\mathsfit}{\encodingdefault}{\sfdefault}{m}{sl}
\SetMathAlphabet{\mathsfit}{bold}{\encodingdefault}{\sfdefault}{bx}{n}

\graphicspath{ {./Figure/} }
\usepackage[font=footnotesize,labelfont=bf]{caption}

\newcommand{\graph}[1]{\mathcal{}}
\providecommand{\keywords}[1]{\textbf{\textit{Keywords: }} #1}

\hypersetup{
bookmarks=true,
bookmarksopen=true,
bookmarksnumbered=true,
unicode=false,
pdftoolbar=true,
pdfmenubar=true,
pdffitwindow=false,
pdfstartview={FitH},
pdftitle={My title},
pdfauthor={Author},
pdfsubject={Subject},
pdfcreator={Creator},
pdfproducer={Producer},
pdfkeywords={keywords},
pdfnewwindow=true,
colorlinks=true,
linkcolor=blue,
citecolor=blue,
filecolor=blue,
urlcolor=blue
}

\begin{document}
\title{\textbf{PhononScore: a phonon-aware scoring function for dynamical stability}}

% \author[1]{Xiao-Qi Han}
% \author[1]{Peng-Jie Guo}
% \author[1,*]{Ze-Feng Gao}
% \author[1,*]{Zhong-Yi Lu} 

\author[1]{Xiao-Qi Han\textcolor{black}{$^\dagger$}}
\author[1,*]{Ze-Feng Gao\textcolor{black}{$^\dagger$}}
% \author[1]{\textcolor{black}{Wen-Kao Li}}
% \author[1]{Peng-Jie Guo}
\author[1,*]{Zhong-Yi Lu}

\affil[1]{\small School of Physics, Renmin University of China, Beijing, China}

\affil[*]{Corresponding authors\vspace{12pt}}

\date{}

\maketitle

\normalsize

\vspace{-28pt} 
\begin{abstract}
{
\small
In recent years, crystal generation models have enabled the design of massive numbers of candidate materials. However, the lack of dynamical stability among generated structures has become a major bottleneck preventing their translation into practical materials discovery. To address this challenge, we propose \textbf{PhononScore}, a phonon-aware scoring function for crystal generation. Unlike computationally expensive explicit phonon calculations, PhononScore directly predicts a unified stability score from crystal structures, enabling rapid ranking of candidate materials according to their dynamical stability with second-level computational cost. PhononScore is built upon periodic atom and line graph representations and jointly optimizes minimum phonon-frequency regression, multi-threshold stability classification, and local geometric stability pattern learning through a multi-task learning framework. A unified ranking objective is further introduced to maximize the enrichment of dynamically stable structures. To train the model, we construct a multi-fidelity phonon dataset containing 157,463 crystal structures, comprising AI-generated crystals, Materials Project materials, and DFT-PBE phonon data, and develop two models, PhononScore and PhononScore-DFT, through a pretraining-and-fine-tuning strategy. On the PhononBench benchmark, PhononScore improves the average dynamical stability rate of candidate pools generated by nine representative crystal generation models from 30.7\% to 83.7\%, achieving a 2.72-fold enrichment of stable structures, while the average stability rate of the Top-10 candidates reaches 97.5\%. On a high-fidelity DFT-PBE phonon benchmark, the DFT-finetuned PhononScore-DFT increases the Top-100 stability rate to 93.0\% and achieves 5–6-fold enrichment of dynamically stable structures under an extremely imbalanced hard-screening scenario. Further analyses demonstrate that PhononScore is not merely a phonon-frequency predictor but a calibrated surrogate scoring function for dynamical stability. Its predicted scores exhibit a monotonic correspondence with the true probability of dynamical stability, faithfully recover nontrivial stability orderings across different stoichiometries, chemically identical systems, and polymorphs, and capture stabilization mechanisms arising from the interplay between local coordination environments and collective lattice dynamics. As a materials-screening tool analogous to scoring functions in drug discovery, PhononScore can serve directly as a dynamical-stability feedback signal for crystal generation, active learning, and reinforcement learning, enabling second-level stability-aware reranking without explicit phonon calculations and providing a unified and efficient dynamical stability evaluator for high-throughput materials discovery, active learning, reinforcement learning, and closed-loop inverse design. We further provide an online PhononScore platform for rapid stability scoring together with a companion web service for phonon spectrum validation.}
\end{abstract}

\vspace{12pt} 

{\footnotesize \textcolor{black}{$^\dagger$These authors contributed equally to this work.}}
\section*{Introduction}\label{sec1}
Recent years have witnessed remarkable advances in crystal generative models, such as MatterGen~\cite{mattergen}, GNoME~\cite{gnome}, InvDesFlow~\cite{InvDesFlow,InvDesFlow-AL,invdesreview}, DiffCSP~\cite{diffcsp}, CrystalFlow~\cite{CrystalFlow}, and CrysVCD~\cite{crysvcd}, which have substantially accelerated inverse materials design and the discovery of novel materials. For example, the generative-design framework InvDesFlow~\cite{invdesreview,InvDesFlow-AL,InvDesFlow} successfully identified the ambient-pressure hydride superconductor Li$_2$AuH$_6$~\cite{liauh}, with a predicted superconducting transition temperature of approximately 140 K, together with a series of superconducting candidates exhibiting transition temperatures beyond the conventional McMillan limit~\cite{prb-bowen}, demonstrating the remarkable potential of artificial intelligence for accelerating the discovery of functional materials. 
Although existing models can produce large numbers of candidates that satisfy compositional and structural constraints, a substantial fraction of these structures are dynamically unstable and therefore unlikely to exist as realizable materials or be experimentally synthesized. A recent large-scale benchmark, PhononBench~\cite{PhononBench,mattersim,mattersim-phonon}, systematically evaluated more than 133,000 structures generated by 7 crystal generation models. Under a stringent dynamical stability criterion requiring the minimum phonon frequency to exceed -0.1 THz, the average stability rate across all generated structures was only 32.15\%. Even the best-performing model, MatterGen, achieved a stability rate of merely 45.05\%, while some large-language-model-based generators exhibited stability rates as low as 14.3\%. These results reveal that a large proportion of structures produced by current crystal generative models remain dynamically unstable, making dynamical stability a critical challenge that limits their practical applicability.

To address this challenge, recent advances in universal machine-learning interatomic potentials (uMLIPs) have provided a promising alternative to expensive first-principles calculations. uMLIPs have evolved from early models such as MEGNet~\cite{MEGNet}, M3GNet~\cite{m3gnet}, CHGNet~\cite{Deng2023CHGNet}, and MACE~\cite{MACE-mp-0} to recent large-scale foundation potentials including MatterSim~\cite{mattersim}, PET-OAM-XL~\cite{pet-oam}, eSEN-30M-OAM~\cite{eSEN-30M-OAM}, and DPA~\cite{dpa2,DPA3,DPA4}, achieving near-DFT accuracy in predicting energies, forces, and stresses. In particular, MatterSim has demonstrated DFT-level reliability for phonon-spectrum prediction and dynamical-stability assessment. However, even the current state-of-the-art models still require minutes per structure for phonon calculations. For example, MatterSim-v1 takes an average of 144 s per structure, while PET-OAM-XL and eSEN-30M-OAM require 272 s and 990 s~\cite{PhononBench}, respectively. Although substantially faster than DFT, such computational costs remain prohibitive for providing online feedback to crystal generative models and for large-scale screening and reranking of millions of candidate structures. As reinforcement learning and generation–evaluation closed-loop frameworks continue to emerge, future crystal generation models will require dynamical-stability evaluation at the second or even sub-second level.

The need for computationally efficient yet reliable dynamical-stability evaluation is not unique to crystal generation. Similar challenges have been extensively addressed in protein structure prediction and drug discovery~\cite{alphafold3}. To efficiently explore vast search spaces, researchers often employ scoring functions as computationally efficient surrogate models for screening~\cite{AutoDock,Eberhardt2021Vina,Flex-jmc}, ranking~\cite{glide,rtmscore}, and optimization, rather than relying exclusively on expensive high-fidelity simulations. Representative examples include docking scores used in virtual screening and confidence metrics such as pLDDT and ipTM in AlphaFold~\cite{AlphaFold-3} for assessing prediction reliability. Although substantially less expensive than the underlying physical calculations, these scoring functions retain strong ranking capability and have become indispensable components of large-scale search and optimization frameworks. The success of this paradigm suggests a similar opportunity for dynamical-stability assessment in crystal generation, where no dedicated scoring function currently exists. A phonon scoring function that can rapidly approximate dynamical stability while preserving reliable ranking performance could directly improve the dynamical stability of generated structures through post-generation reranking and provide real-time feedback during the generation process.

In this work, we introduce \textbf{PhononScore}, the first phonon-aware scoring function for crystal generation, and reformulate dynamical-stability assessment as a \emph{scoring-function learning} problem rather than explicit phonon prediction. Instead of performing computationally expensive phonon calculations, PhononScore directly predicts a unified stability score that serves as a surrogate objective for dynamical stability while preserving the relative stability ordering among candidate structures. To this end, we construct a large-scale multi-fidelity phonon dataset containing AI-generated crystals, Materials Project materials, and DFT-PBE phonon calculations, and develop a multi-task learning framework that jointly optimizes minimum phonon-frequency regression, multi-threshold stability classification, local geometric stability learning, and a ranking objective tailored for stable-structure enrichment. Based on a pretraining-and-fine-tuning strategy, we further develop two complementary models, \textbf{PhononScore} and \textbf{PhononScore-DFT}, for efficient large-scale screening and high-fidelity stability assessment, respectively. Extensive experiments on the PhononBench benchmark demonstrate that PhononScore increases the average dynamical-stability rate of candidates generated by nine representative crystal generation models from \textbf{30.7\% to 83.7\%}, achieving \textbf{2.72$\times$ enrichment} of dynamically stable structures. On the high-fidelity DFT-PBE benchmark, PhononScore-DFT further achieves a \textbf{93.0\% Top-100 stability rate} and \textbf{5--6$\times$ enrichment} of dynamically stable structures under extremely imbalanced hard-screening settings. Beyond predictive accuracy, we show that PhononScore behaves as a calibrated surrogate scoring function whose outputs exhibit a monotonic correspondence with the true probability of dynamical stability, faithfully recover nontrivial stability orderings across different stoichiometries, chemically related systems, and polymorphs, and capture stabilization mechanisms arising from the interplay between local coordination environments and collective lattice dynamics. By enabling efficient post-generation reranking and providing real-time feedback during optimization, PhononScore establishes a practical foundation for high-throughput crystal generation, active learning, reinforcement learning, and closed-loop materials discovery.

\section*{Results}\label{sec-results}

\subsection*{PhononScore Framework}
\begin{figure*}[t!]
		\centering  
		\includegraphics[width=1.0\linewidth]{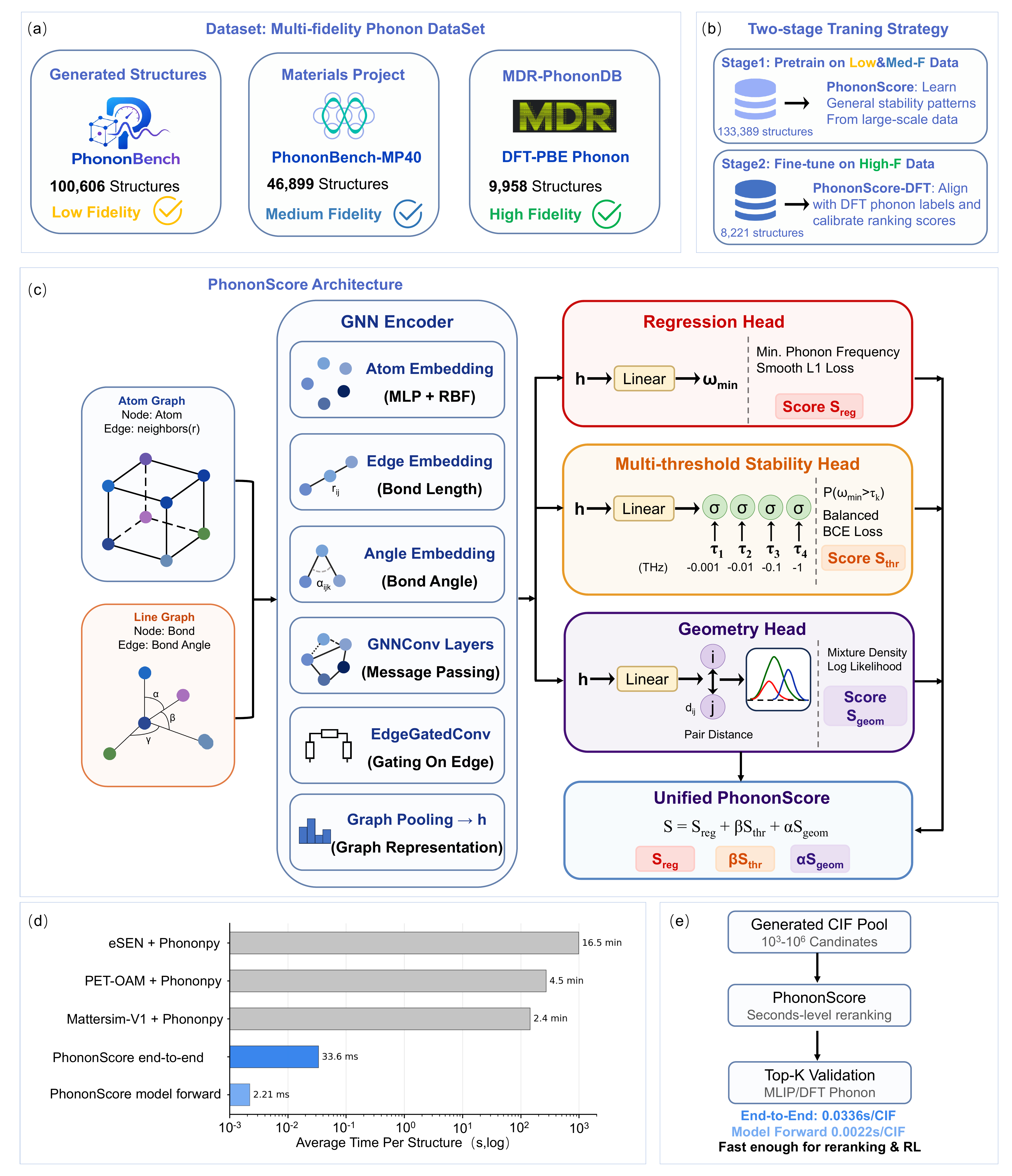}
		\caption{\textbf{Overview of the PhononScore framework.}
(a) Multi-fidelity phonon dataset used for training.
(b) Two-stage training strategy consisting of large-scale pretraining followed by DFT fine-tuning.
(c) PhononScore architecture. A shared graph neural network predicts the minimum phonon frequency, multi-threshold stability, and local geometry likelihood, which are combined into a unified ranking score.
(d) Comparison of inference efficiency between PhononScore and conventional phonon calculations.
(e) Deployment workflow, where PhononScore rapidly reranks generated candidates before first-principles phonon validation.}
		\label{phononscore-fram} 
\end{figure*}

To enable efficient assessment of dynamical stability in generated crystal structures, we introduce PhononScore (Fig.~\ref{phononscore-fram}), a phonon-aware scoring function designed for crystal generation. Unlike conventional phonon calculations that explicitly solve lattice vibrations and compute full phonon spectra, PhononScore directly predicts a continuous ranking score from crystal structures, providing a fast estimate of their stability potential. Higher scores generally indicate greater dynamical stability. Rather than minimizing phonon-frequency prediction errors or reconstructing complete phonon spectra, PhononScore is specifically designed to optimize ranking performance and stable-structure enrichment in high-throughput screening. Its design philosophy is analogous to scoring functions used in drug discovery, where low-cost evaluations are employed to rapidly prioritize the most promising candidates while reducing the search space for subsequent high-fidelity calculations.

As shown in Fig.~\ref{phononscore-fram}c, PhononScore represents each input crystal structure as a periodic atom graph and its corresponding line graph, jointly encoding atomic identities, bond lengths, bond angles, and local coordination environments. This graph representation allows the model to capture structural features associated with dynamical stability across diverse crystal systems. Because dynamical stability involves not only continuous variations in phonon frequencies but also stability-threshold classification and local structural plausibility, PhononScore adopts a multi-task learning framework. Specifically, the model simultaneously learns minimum phonon-frequency prediction, multi-threshold dynamical-stability classification, and local geometric patterns characteristic of stable structures, and integrates these complementary signals into a unified ranking score. This design enables PhononScore to preserve continuous ranking capability while improving the identification and enrichment of dynamically stable structures.

To train and evaluate PhononScore, we constructed a multi-fidelity phonon dataset containing 157,463 crystal structures (Fig.~\ref{phononscore-fram}a), including 100,606 generated structures~\cite{PhononBench}, 46,899 Materials Project structures with fewer than 40 atoms~\cite{PhononBench,PhononBench-MP40}, and 9,958 structures with DFT-PBE phonon labels~\cite{mattersim-phonon,MDRPhononDB}. The first two datasets were labeled using minimum phonon frequencies obtained from MatterSim coupled with phonopy and served as the primary source for large-scale pretraining. Among them, the generated structures most closely match the distribution of candidates produced by crystal generation models, while the Materials Project structures provide additional structural and chemical priors. Together with the DFT-PBE phonon dataset, these data sources establish a low-, medium-, and high-fidelity hierarchy for developing and evaluating PhononScore.
To prevent data leakage and ensure rigorous evaluation, we adopted a formula-level data splitting strategy, in which all reduced formulas appearing in the test sets were completely excluded from the training data. The final benchmark contains 10,000 structures covering generated-structure reranking, transferability to DFT phonon labels, and out-of-distribution evaluation, corresponding to 9,818 unique reduced formulas. After formula-level filtering, the pretraining set contains 133,389 MatterSim-labeled structures, while the fine-tuning set contains 8,221 DFT-PBE phonon samples. This protocol enforces strict formula-level separation between training and test data, thereby minimizing the risk of data leakage arising from compositional overlap. 

As shown in Fig.~\ref{phononscore-fram}b, model training follows a two-stage strategy. PhononScore is first pretrained on large-scale MatterSim-derived phonon labels to learn general stability patterns from both generated structures and experimentally known materials. The pretrained model is then further fine-tuned and calibrated using high-fidelity DFT-PBE phonon data, resulting in PhononScore-DFT, which is more closely aligned with first-principles phonon calculations. This strategy combines the scalability of MatterSim-derived labels with the physical fidelity of DFT phonons, yielding a ranking function that is both data-efficient and physically grounded.

As shown in Fig.~\ref{phononscore-fram}d, PhononScore achieves orders-of-magnitude higher computational efficiency than conventional phonon calculations while preserving its ability to prioritize dynamically stable structures. This efficiency makes it suitable not only for large-scale crystal screening but also for reinforcement learning and other closed-loop materials discovery frameworks. In practical applications (Fig.~\ref{phononscore-fram}e), PhononScore serves as a model-agnostic post-processing module that rapidly scores and reranks candidate structures generated by any crystal generation model, allowing computationally expensive phonon calculations to be focused on the most promising candidates. We next systematically evaluate its reranking performance and stable-structure enrichment capability on large-scale candidate pools generated by multiple crystal generation models.

\subsection*{PhononScore enables efficient enrichment of dynamically stable structures}

To evaluate the ability of PhononScore as a dynamical-stability scoring function for crystal generation, we adopted PhononBench~\cite{PhononBench} as the benchmark. PhononBench systematically assesses the dynamical stability of large-scale candidate structures generated by multiple state-of-the-art crystal generative models and represents one of the largest benchmarks currently available in this area. Based on PhononBench, we further constructed a reranking benchmark to directly evaluate the capability of PhononScore to enrich dynamically stable structures. The benchmark consists of 9,000 candidate structures collected from nine crystal generative models, with 1,000 randomly sampled structures from each source. Following the PhononBench protocol, the ground-truth label for each structure is defined by its minimum phonon frequency computed using MatterSim~\cite{mattersim}. During evaluation, PhononScore is used solely for scoring and ranking candidate structures, while all stability assessments are performed using the corresponding ground-truth minimum phonon frequencies. This setup enables a direct and unbiased evaluation of its ability to prioritize and enrich dynamically stable structures. Unless otherwise specified, a structure is considered dynamically stable when its minimum phonon frequency satisfies $\omega_{\mathrm{min}} > -0.1$ THz.

\begin{figure*}[t!]
		\centering  
		\includegraphics[width=1.0\linewidth]{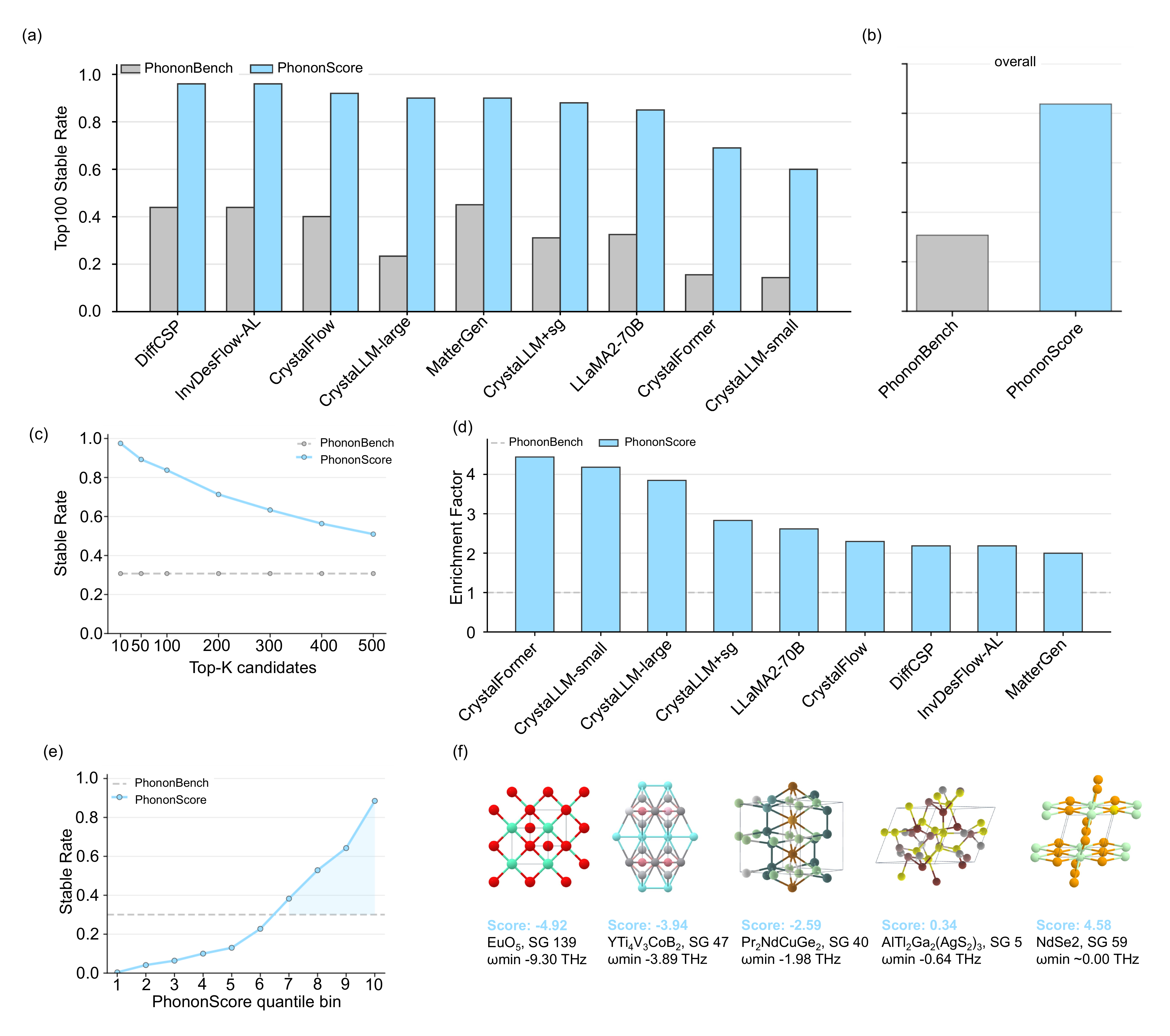}
		\caption{\textbf{PhononScore enables efficient enrichment of dynamically stable structures in crystal-generation candidate pools.}
(a) Dynamical stability rates of the original candidate pools (PhononBench) and the Top-100 candidates selected by PhononScore across nine crystal generation models. 
(b) Average dynamical stability rates before and after reranking. The average stability rate increases from 30.7\% in the original candidate pools to 83.7\% in the Top-100 candidates selected by PhononScore.
(c) Average dynamical stability rate as a function of the Top-$K$ cutoff, evaluated on candidate pools containing 1,000 structures per generation source.
(d) Enrichment factors achieved by PhononScore across different generation sources, defined as the ratio between the Top-100 stability rate after reranking and the stability rate of the original candidate pool.
(e) Fraction of dynamically stable structures as a function of PhononScore. All 9,000 candidate structures are divided into equal-sized bins according to their PhononScore values, and the stability rate is calculated within each bin.
(f) Representative crystal structures spanning different PhononScore ranges. The corresponding minimum phonon frequencies progressively evolve from strongly imaginary-frequency regimes toward dynamically stable regions as the score increases.}
		\label{Benchmark-main} 
\end{figure*}

As shown in Fig.~\ref{Benchmark-main}a, PhononScore consistently enriches dynamically stable structures across all generation sources. After reranking with PhononScore, the Top-100 candidates achieve substantially higher dynamical stability rates than their original candidate pools. Across multiple crystal generation models, including DiffCSP, InvDesFlow-AL, CrystalFlow, MatterGen, CrystalLLM, CrystalFormer, and LLaMA2-70B, the Top-100 stability rates exceed 60\%, reaching 96\%, 96\%, and 92\% for DiffCSP, InvDesFlow-AL, and CrystalFlow, respectively. Even for CrystalFormer and CrystalLLM-small, which exhibit relatively low stability rates in their original candidate pools, PhononScore increases the Top-100 stability rates by 45 and 53 percentage points, respectively, demonstrating its ability to identify dynamically stable structures from candidate pools dominated by unstable materials. As summarized in Fig.~\ref{Benchmark-main}b, the average dynamical stability rate across the original candidate pools from the nine generation sources is only 30.7\%, whereas reranking with PhononScore raises the average Top-100 stability rate to 83.7\%, corresponding to a 2.72-fold enrichment of dynamically stable structures. These results indicate that PhononScore captures transferable ranking signals associated with true dynamical stability rather than source-specific statistical biases, enabling robust generalization across diverse crystal generation models.

Fig.~\ref{Benchmark-main}c further evaluates the screening efficiency of PhononScore under different Top-$K$ cutoffs, with the candidate pool size fixed at 1,000 structures. The strongest enrichment is achieved among the highest-ranked candidates, where the average dynamical stability rates reach 97.50\%, 89.25\%, and 83.75\% for Top-10, Top-50, and Top-100 structures, respectively. Although the stability rate gradually decreases as $K$ increases, it remains as high as 50.95\% at Top-500, substantially exceeding the 30.74\% average stability rate of the original PhononBench candidate pool.

Fig.~\ref{Benchmark-main}d quantifies the improvement achieved by PhononScore from the perspective of enrichment. Here, the enrichment factor is defined as the ratio between the Top-100 stability rate after reranking with PhononScore and the stability rate of the original PhononBench candidate pool. All generation sources exhibit substantial enrichment, ranging from 2.00$\times$ to 4.44$\times$. In particular, CrystalFormer-mp20 and CrystalLLM-gen achieve enrichment factors of 4.44$\times$ and 4.18$\times$, respectively, indicating that PhononScore is especially effective for candidate pools with low initial stability rates. This behavior is well aligned with the intended role of a scoring function: when a generative model produces large numbers of dynamically unstable structures, PhononScore can substantially reduce the computational cost of downstream phonon validation by prioritizing candidates with a higher likelihood of dynamical stability.

To further examine whether PhononScore reflects true dynamical stability rather than merely serving as a heuristic screening metric, we divided all 9,000 candidate structures into equal-sized bins according to their PhononScore values and calculated the fraction of dynamically stable structures within each bin (Fig.~\ref{Benchmark-main}e). A clear monotonic relationship is observed, with the stability probability increasing from below 1\% in the lowest-score bin to nearly 90\% in the highest-score bin. This well-calibrated trend indicates that PhononScore captures meaningful information associated with dynamical stability. Beyond Top-$K$ screening, it can therefore serve as a continuous stability signal for candidate prioritization and provide reliable feedback for active learning and reinforcement-learning-based optimization.
This ranking trend is further supported at the level of individual crystal structures. Representative examples in Fig.~\ref{Benchmark-main}f show that, as the PhononScore increases, the corresponding minimum phonon frequency gradually evolves from strongly imaginary-frequency regimes toward and eventually beyond the dynamical-stability threshold. Low-scoring structures are typically characterized by pronounced soft modes and large imaginary frequencies, whereas high-scoring structures are more likely to reside in locally stable regions of the potential-energy surface. These representative examples provide an intuitive structural interpretation of the ranking behavior learned by PhononScore and demonstrate that the ranking signal is physically meaningful and closely associated with the underlying dynamical stability of crystal structures.

Collectively, these results demonstrate that PhononScore is not merely a ranking heuristic or a predictor of minimum phonon frequency, but a calibrated surrogate score for dynamical stability. By rapidly enriching stable structures while preserving meaningful stability rankings across diverse crystal generation models, PhononScore provides an effective scoring-function framework for large-scale crystal generation, candidate reranking, and closed-loop materials discovery.
These results establish PhononScore as a practical analogue of scoring functions widely used in drug discovery, enabling rapid stability-aware ranking of generated crystal structures without explicit phonon calculations.

To determine whether the performance gains of PhononScore arise merely from direct prediction of the minimum phonon frequency, we further compared it with an ALIGNN baseline trained to regress $\omega_{\min}$ (Table~\ref{tab:ablation_alignn}). Although direct $\omega_{\min}$ prediction already improves the average dynamical stability rate of the Top-100 candidates from 30.7\% to 69.4\%, PhononScore consistently achieves superior performance across all generation sources, further increasing the average stability rate to 83.8\%. These results suggest that PhononScore learns a stability-oriented ranking representation beyond direct $\omega_{\min}$ regression, making it more effective for prioritizing dynamically stable crystal structures. Detailed comparisons under different stability criteria are provided in the Appendix.

\begin{table}[H]
\centering
\small
\caption{
\textbf{Comparison between direct phonon-frequency prediction and PhononScore-based reranking.}
Values represent the true dynamical stability rate (\%) of the Top-100 structures selected from 1,000 held-out candidates for each generator. Dynamical stability is defined as $\omega_{\min}>-0.1$ THz. Improvement denotes the increase in Top-100 stability rate achieved by PhononScore over ALIGNN.
}
\label{tab:ablation_alignn}
\begin{tabular}{lcccc}
\toprule
Generator Source &
PhononBench (\%) &
ALIGNN ($\omega_{\min}$) (\%) &
PhononScore (\%) &
Improvement (\%)\\
\midrule
DiffCSP             & 43.9 & 72.0 & \textbf{96.0} & +24.0 \\
InvDesFlow-AL             & 43.7 & 72.0 & \textbf{96.0} & +24.0 \\
CrystalFlow         & 40.1 & 74.0 & \textbf{92.0} & +18.0 \\
CrystalLLM-large    & 23.4 & 73.0 & \textbf{90.0} & +17.0 \\
MatterGen           & 45.1 & 73.0 & \textbf{90.0} & +17.0 \\
CrystalLLM+sg       & 31.1 & 73.0 & \textbf{88.0} & +15.0 \\
LLaMA2-70B          & 32.5 & 77.0 & \textbf{85.0} & +8.0 \\
CrystalFormer       & 15.5 & 63.0 & \textbf{69.0} & +6.0 \\
CrystalLLM-small    & 14.3 & 50.0 & \textbf{60.0} & +10.0 \\
\midrule
Overall             & 30.7 & 69.4 & \textbf{83.8} & +14.4 \\
\bottomrule
\end{tabular}
\end{table}

\subsection*{Transferability of PhononScore to high-fidelity DFT phonon labels}
\begin{figure*}[t!]
		\centering  
		\includegraphics[width=1.0\linewidth]{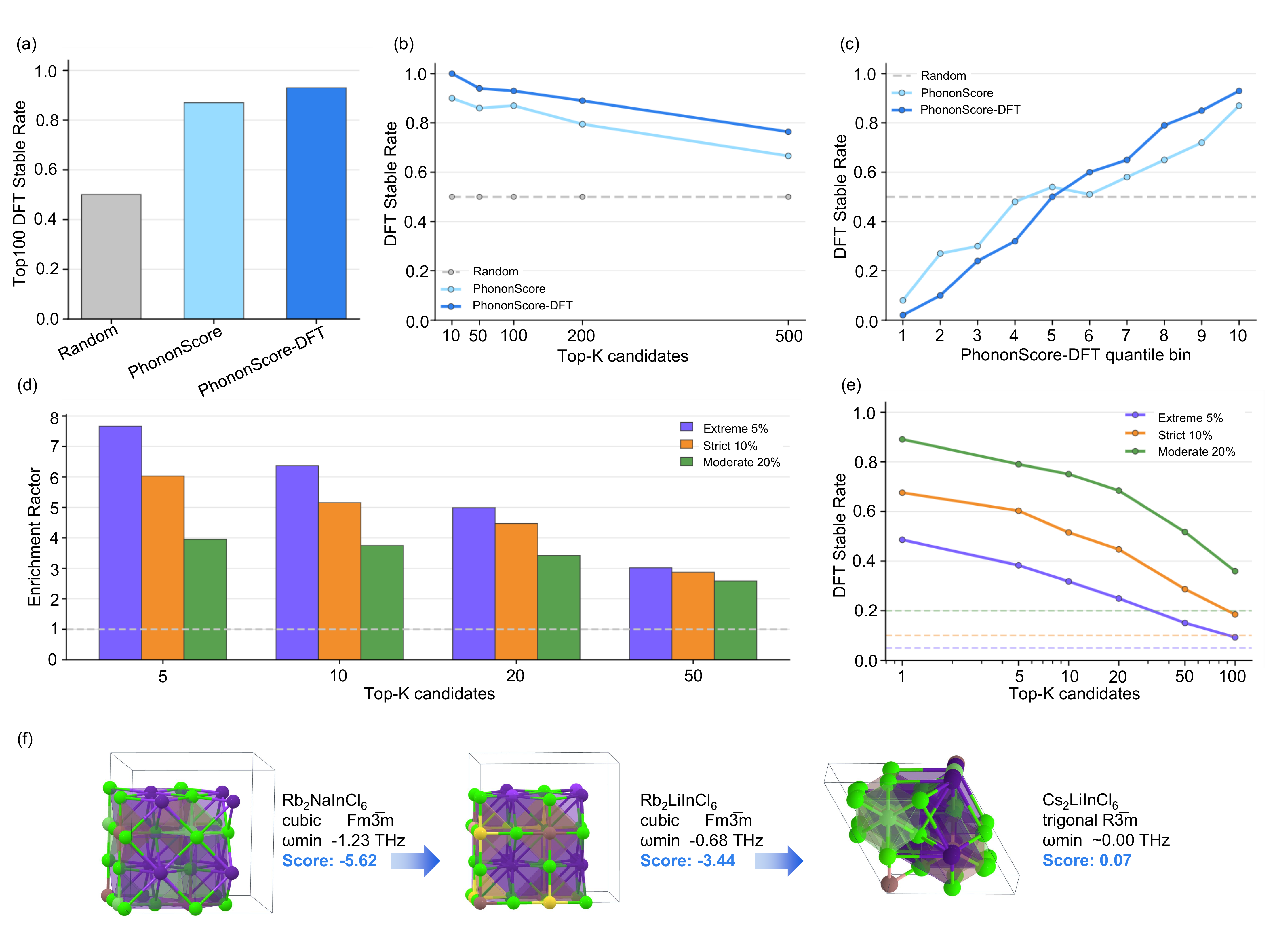}
		\caption{\textbf{PhononScore transfers effectively to high-fidelity DFT phonon labels and enables robust hard-screening of stable structures.}
(a) Dynamical stability rates of the Top-100 candidates selected by the pretrained PhononScore and the DFT-finetuned PhononScore-DFT on a balanced DFT-PBE test set containing 1,000 Materials Project structures. The random baseline stability rate is 50\%.
(b) DFT dynamical stability rates under different Top-$K$ cutoffs for PhononScore and PhononScore-DFT. PhononScore-DFT achieves stability rates of 100.0\%, 94.0\%, and 93.0\% for the Top-10, Top-50, and Top-100 candidates, respectively.
(c) Fraction of DFT-stable structures as a function of PhononScore-DFT. All test structures are divided into ten equal-sized bins according to their scores, and the stability rate is calculated within each bin. The stability probability increases monotonically from 2.0\% to 93.0\% across the score range.
(d) Top-10 stability rates under DFT hard-screening benchmarks with different fractions of stable structures in the candidate pool (Extreme: 5\%, Strict: 10\%, Moderate: 20\%). Results are averaged over 1,000 random resampling trials.
(e) Screening performance as a function of Top-$K$ under the Strict 10\% hard-screening setting. The average number and fraction of DFT-stable structures recovered among the highest-ranked candidates are reported.
(f) Representative A$_2$BInCl$_6$ halide perovskite-derived structures illustrating the relationship between PhononScore-DFT and local stabilization mechanisms. As octahedral size mismatch is reduced and octahedral tilting stabilizes the framework, the DFT minimum phonon frequency increases from $-1.225$ THz to nearly $0$ THz, accompanied by a corresponding increase in PhononScore-DFT from $-5.615$ to $0.073$.
}
		\label{Benchmark-DFT} 
\end{figure*}

To assess whether PhononScore can transfer to higher-fidelity DFT phonon labels, we evaluated it on a balanced DFT-PBE test set containing 1,000 Materials Project structures. Among them, 500 structures satisfy the dynamical-stability criterion of $\omega_{\min}>-0.1$ THz, while the remaining 500 are dynamically unstable, resulting in a random-baseline stability rate of 50\%. Two scoring models were compared: PhononScore, which was pretrained on MatterSim-derived phonon labels, and PhononScore-DFT, which was further fine-tuned on DFT-PBE phonon samples starting from the pretrained model. As shown in Fig.~\ref{Benchmark-DFT}a, the pretrained PhononScore already exhibits strong enrichment of DFT-stable structures, increasing the true stability rate of the Top-100 candidates from the random baseline of 50\% to 87.0\%. After DFT-PBE fine-tuning, PhononScore-DFT further improves the Top-100 stability rate to 93.0\%. As shown in Fig.~\ref{Benchmark-DFT}b, under more stringent Top-$K$ selection, PhononScore-DFT achieves stability rates of 100.0\%, 94.0\%, and 93.0\% for the Top-10, Top-50, and Top-100 candidates, respectively, demonstrating its strongest enrichment capability at the top of the ranking and making it particularly suitable for prioritizing candidates before expensive phonon calculations. Beyond Top-$K$ metrics, we further divided all DFT-PBE test structures into ten equal-sized bins according to their PhononScore-DFT values, from low to high. As shown in Fig.~\ref{Benchmark-DFT}c, the true DFT stability rate increases monotonically from 2.0\% in the lowest-score bin to 93.0\% in the highest-score bin, indicating that the score is not merely a heuristic for Top-$K$ selection but also provides a meaningful continuous ranking of stability likelihood. Consistently, the Spearman rank correlation between PhononScore-DFT and the DFT minimum phonon frequency reaches 0.596, exceeding the value of 0.527 obtained by the pretrained model, demonstrating that DFT fine-tuning further calibrates the model's ranking with respect to DFT phonon labels.

To mimic realistic high-throughput discovery scenarios where dynamically stable structures are rare, we further constructed a DFT hard-screening benchmark. Instead of rerunning inference, candidate pools were repeatedly resampled from the balanced DFT-PBE test set by combining a small number of stable structures with a large number of unstable structures. Three difficulty levels were considered, corresponding to candidate pools with stable fractions of 5\% (Extreme), 10\% (Strict), and 20\% (Moderate), thereby simulating increasingly imbalanced screening environments. 
As shown in Fig.~\ref{Benchmark-DFT}d, PhononScore-DFT consistently enriches DFT-stable structures across all hard-screening settings. In the Strict 10\% setting, each candidate pool contains 20 stable and 180 unstable structures and is repeated over 1000 random resampling trials. PhononScore-DFT increases the Top-10 stable rate from the random baseline of 9.9\% to 51.5\%, corresponding to a 5.16-fold enrichment. Even under the more challenging Extreme 5\% setting, the Top-10 stable rate remains 31.8\%, representing a 6.36-fold enrichment over the pool baseline. As the stable fraction increases from 5\% to 20\%, the Top-10 stable rate further rises from 31.8\% to 75.0\%.
Fig~\ref{Benchmark-DFT}e further examines the dependence on screening depth. Under the Strict 10\% setting, the average number of DFT-stable structures recovered within the Top-20 candidates increases from 1.99 under random selection to 8.95 using PhononScore-DFT, while the Top-20 and Top-50 stable rates reach 44.7\% and 28.7\%, respectively. These results demonstrate that PhononScore-DFT remains highly effective even when stable structures are scarce, substantially increasing the concentration of DFT-stable candidates at the top of the ranked list. 

As a mechanistic example, we examine an A$_2$BInCl$_6$ halide double-perovskite-derived series. Rb$_2$NaInCl$_6$ and Rb$_2$LiInCl$_6$ crystallize in the cubic Fm$\bar{3}$m structure with untilted corner-sharing octahedra, whereas Cs$_2$LiInCl$_6$ adopts a distorted trigonal R$\bar{3}$m framework featuring octahedral tilting. Such symmetry lowering provides an effective structural pathway for releasing soft-mode instabilities present in the high-symmetry cubic phase. Consistent with this stabilization trend, the DFT-PBE minimum phonon frequency increases from $-1.225$ THz to nearly $0$ THz, while the corresponding PhononScore-DFT rises from $-5.615$ to $0.073$. This case demonstrates that PhononScore captures physically meaningful local stabilization mechanisms associated with dynamical stability, rather than merely reflecting composition-level statistical priors.

\subsection*{PhononScore-DFT correctly recovers stability ordering in the K--I--O system}

While the A$_2$BInCl$_6$ example demonstrates that PhononScore-DFT is sensitive to local stabilization mechanisms, the compounds differ in chemical composition, making stability trends partially correlated with compositional changes. A more challenging test is whether the model can distinguish dynamical stability within a single chemical system, where the elemental constituents remain the same and stability differences arise primarily from stoichiometric variation, changes in local coordination environments, and collective lattice-dynamical effects rather than composition alone.

\begin{figure*}[t!]
		\centering  
		\includegraphics[width=1.0\linewidth]{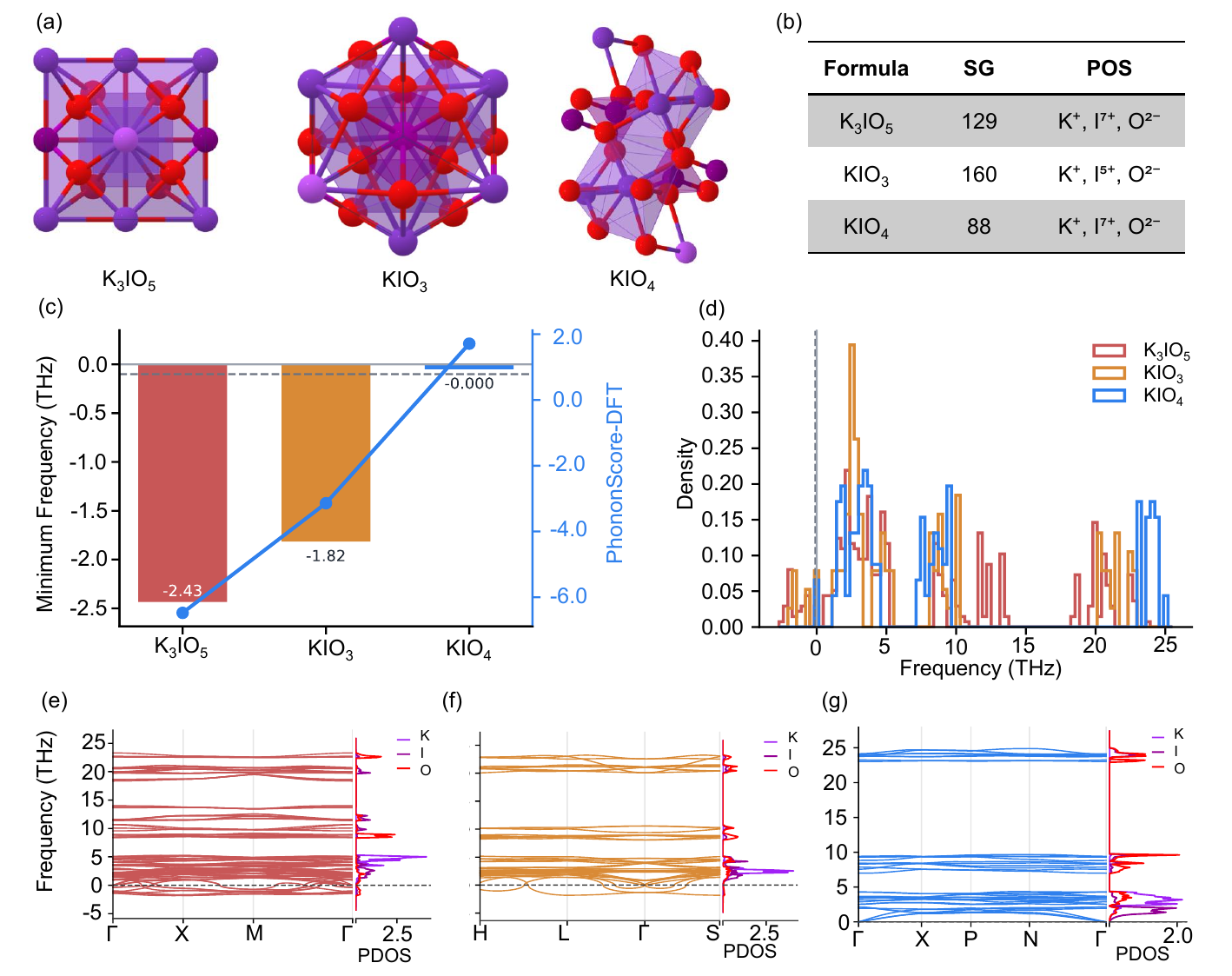}
\caption{
\textbf{PhononScore-DFT Correctly Recovers Stability Ordering in the K--I--O System.}
(a) Crystal structures of KIO$_3$, K$_3$IO$_5$, and KIO$_4$, illustrating the structural evolution with increasing oxygen content.
(b) Chemical compositions, space groups, and iodine oxidation states of the three compounds. KIO$_3$ contains I$^{5+}$ in the R3m structure, whereas K$_3$IO$_5$ and KIO$_4$ contain I$^{7+}$ in the P4/nmm and I4$_1$/a structures, respectively.
(c) Comparison of the DFT minimum phonon frequency ($\omega_{\min}$) and PhononScore-DFT for the three compounds. PhononScore-DFT reproduces the same stability ordering as DFT phonon calculations, namely K$_3$IO$_5$ $<$ KIO$_3$ $<$ KIO$_4$.
(d) Distributions of phonon frequencies for KIO$_3$, K$_3$IO$_5$, and KIO$_4$. The negative-frequency region progressively disappears as the system evolves toward the dynamically stable KIO$_4$ phase.
(e) Phonon band structure and projected phonon density of states (PDOS) of K$_3$IO$_5$. Imaginary phonon modes extend over a broad range of wave vectors, indicating collective lattice instability. The imaginary-frequency region is dominated by O vibrations, with additional contributions from K atoms.
(f) Phonon band structure and PDOS of KIO$_3$. The instability is concentrated in a smaller number of low-frequency soft modes. O atoms contribute approximately 87\% of the imaginary-frequency spectral weight, indicating that the instability primarily originates from distortions of the oxygen coordination framework.
(g) Phonon band structure and PDOS of KIO$_4$. No imaginary phonon modes are observed. The regular IO$_4$ tetrahedral units and high-symmetry crystal framework suppress oxygen-related soft modes and result in a dynamically stable structure. Together, this K--I--O series demonstrates that dynamical stability is governed by collective lattice vibrations rather than local chemical descriptors alone, while PhononScore-DFT successfully recovers the corresponding nontrivial stability ordering.
}
		\label{eg-KIO} 
\end{figure*}

To investigate this question, we selected three representative compounds from the K--I--O system, KIO$_3$, K$_3$IO$_5$, and KIO$_4$, for comparative analysis (Fig.~\ref{eg-KIO}a). As summarized in Fig.~\ref{eg-KIO}b, these compounds exhibit distinct crystal structures and iodine oxidation states. KIO$_3$ crystallizes in the R3m space group, where iodine adopts a +5 oxidation state and exhibits a highly asymmetric I--O coordination environment, characterized by both short ($\sim$1.82~\AA) and long ($\sim$2.65~\AA) I--O bonds. Upon increasing oxygen content, the structure evolves into tetragonal K$_3$IO$_5$ (P4/nmm), in which iodine is oxidized to +7 and forms distorted IO$_5$ square pyramids with I--O bond lengths ranging from 1.80 to 1.85~\AA. Further oxidation yields KIO$_4$, which crystallizes in the high-symmetry I4$_1$/a space group and consists of nearly regular IO$_4$ tetrahedra with uniform I--O bond lengths of approximately 1.77--1.78~\AA.

DFT phonon calculations reveal markedly different dynamical stabilities among these compounds. As shown in Fig.~\ref{eg-KIO}c, the minimum phonon frequencies of K$_3$IO$_5$, KIO$_3$, and KIO$_4$ are $-2.434$, $-1.817$, and $0.000$~THz, respectively, while their corresponding PhononScore-DFT values are $-6.475$, $-3.140$, and $1.712$. Notably, despite possessing a higher iodine oxidation state and shorter I--O bonds, K$_3$IO$_5$ is dynamically less stable than KIO$_3$, whereas KIO$_4$ becomes fully stable. PhononScore-DFT reproduces exactly the same stability ordering predicted by DFT phonon calculations, namely K$_3$IO$_5$ $<$ KIO$_3$ $<$ KIO$_4$, demonstrating its ability to capture stability variations induced by stoichiometric changes within a chemically consistent system. The overall phonon-frequency distributions shown in Fig.~\ref{eg-KIO}d further support this trend: the negative-frequency region gradually disappears as the system evolves from K$_3$IO$_5$ and KIO$_3$ to KIO$_4$, indicating a progressive enhancement of dynamical stability.

To further understand the physical origin of this ordering, we analyzed the phonon band structures and projected phonon density of states (PDOS) shown in Fig.~\ref{eg-KIO}e--g. For K$_3$IO$_5$ (Fig.~\ref{eg-KIO}e), imaginary phonon modes persist over a broad range of wave vectors, with multiple low-frequency branches entering the negative-frequency region. This behavior indicates a genuine collective lattice instability rather than numerical noise. PDOS analysis shows that approximately 74\% of the imaginary-frequency contribution originates from O atoms, accompanied by low-frequency displacements of K atoms, while the contribution from I atoms remains relatively small. These results suggest that the instability primarily arises from cooperative distortions of the K--O/I--O polyhedral framework.
In contrast, the imaginary modes in KIO$_3$ (Fig.~\ref{eg-KIO}f) are concentrated within a limited number of low-frequency soft branches, indicating a more localized instability. The PDOS reveals that approximately 87\% of the imaginary-frequency contribution originates from O atoms, whereas I and K contribute only about 8\% and 5\%, respectively. Therefore, the dynamical instability of KIO$_3$ is likewise dominated by soft vibrational modes associated with the oxygen framework rather than by local I--O bond vibrations.
For KIO$_4$ (Fig.~\ref{eg-KIO}g), no imaginary phonon modes are observed. The combination of regular IO$_4$ tetrahedra, a high-symmetry crystal framework, and a uniform local bonding environment effectively suppresses oxygen-related soft modes, resulting in a dynamically stable structure. Collectively, this K--I--O series demonstrates that dynamical stability cannot be inferred solely from oxidation state or local bond length. Instead, it emerges from the interplay among local coordination environments, structural complexity, and collective lattice vibrations. The ability of PhononScore-DFT to accurately recover this nontrivial stability ordering indicates that the model has learned structural descriptors closely related to dynamical stability rather than relying on simple chemical heuristics.

Furthermore, PhononScore-DFT consistently recovers the stability ordering predicted by DFT minimum phonon frequencies across a variety of scenarios, including polymorphic systems, stoichiometric variations, and element-family substitutions. These results indicate that the model captures key structural features associated with dynamical stability and maintains robust ranking performance across diverse materials classes. Representative case studies are provided in the Appendix.

\section*{Discussion}\label{sec12}
Recent advances in crystal generation models have enabled the creation of millions of candidate structures, shifting the primary bottleneck in materials discovery from generation to screening. Although first-principles phonon calculations provide accurate assessments of dynamical stability, their computational cost precludes large-scale evaluation of generated candidates. PhononScore addresses this challenge by formulating dynamical stability assessment as a ranking problem rather than a conventional regression task. By jointly learning minimum phonon frequencies, multi-threshold stability labels, and local geometric patterns, together with a ranking objective that directly optimizes stable-structure enrichment, PhononScore aligns its optimization target with the practical objective of high-throughput crystal screening. Our results demonstrate that directly optimizing the ranking of stable structures is more effective for materials discovery than further improving phonon-frequency prediction accuracy, providing an efficient bridge between crystal generation and high-fidelity phonon calculations.

Despite its strong performance across multiple crystal generation models and multi-fidelity datasets, PhononScore remains fundamentally limited by the quality of available supervision and the approximations underlying current phonon calculations. The model is trained primarily on minimum phonon frequencies obtained within the harmonic approximation, whereas publicly available data for strongly anharmonic materials, strongly correlated systems, and materials with pronounced electron--phonon coupling remain limited in both accuracy and scale. These limitations may restrict the generalizability of the learned scoring function to more complex materials systems. Future advances in high-fidelity DFT calculations, experimental phonon measurements, and electronic structure theory are expected to further improve the robustness and transferability of PhononScore.

As a model-agnostic dynamical stability scoring function, PhononScore can be readily integrated into existing AI-driven materials discovery workflows. It can serve as a fast reranking module for crystal generation, a reward function for reinforcement learning, an acquisition strategy for active learning, and a dynamical stability evaluator in closed-loop discovery pipelines. Analogous to scoring functions in virtual screening for drug discovery, we anticipate that fast dynamical stability scoring will become a fundamental component of future AI-driven materials discovery.

\section*{Methods}
\label{sec-method}

\subsection*{Crystal graph representation}
PhononScore takes crystal structures in CIF format as input and represents each crystal as a periodic atom graph together with its corresponding line graph, following previous graph representations for crystalline materials~\cite{Choudhary2021,DPA3,DimeNet,CrysToGraph}. In the atom graph, atoms are treated as nodes and periodic neighbor bonds as edges, whereas the line graph explicitly models bond-angle interactions between neighboring bonds. This representation jointly encodes atomic identities, bond lengths, bond angles, and local coordination environments. The resulting graph representation is processed by a shared graph neural network encoder through iterative message passing on both the atom graph and line graph, producing a crystal-level representation

\begin{equation}
\mathbf{h}=f_{\mathrm{GNN}}(G,L(G)),
\end{equation}
where $G$ and $L(G)$ denote the periodic atom graph and its corresponding line graph, respectively. The shared representation $\mathbf{h}$ is subsequently used by all downstream prediction branches.

\subsection*{Multi-task learning for dynamical stability}
Unlike conventional phonon prediction models that optimize only minimum phonon-frequency regression, PhononScore formulates dynamical-stability assessment as a multi-task learning problem. Dynamical stability is characterized not only by the continuous value of the minimum phonon frequency, but also by discrete stability boundaries and physically plausible local coordination environments. Accordingly, PhononScore jointly optimizes three complementary objectives: minimum phonon-frequency regression, multi-threshold dynamical-stability classification, and local geometry likelihood estimation. The regression branch predicts the minimum phonon frequency through a linear projection,
\begin{equation}
\hat{\omega}_{\min}=W_r\mathbf{h}+b_r,
\end{equation}
which is supervised using a Smooth L1 regression loss that is less sensitive to strongly unstable structures with large imaginary phonon frequencies.To explicitly capture stability transitions, multiple dynamical-stability thresholds are introduced,
\begin{equation}
\tau\in
\{-0.001,-0.01,-0.1,-1.0\}\ {\rm THz},
\end{equation}
which correspond to progressively stricter stability criteria. Binary labels are then constructed as
\begin{equation}
y_k=\mathbb{I}(\omega_{\min}>\tau_k),
\end{equation}
where $\mathbb{I}(\cdot)$ denotes the indicator function. The classification branch predicts the probability that each stability criterion is satisfied,
\begin{equation}
p_k=\sigma(W_k\mathbf{h}+b_k),
\end{equation}
and is optimized using class-balanced binary cross entropy. To preserve the ordinal relationship among different stability levels, an additional monotonicity constraint is introduced to penalize inconsistent predictions across neighboring thresholds.

\subsection*{Geometry-aware stability modeling}
Besides predicting phonon frequencies, PhononScore incorporates a geometry branch based on a mixture density network (MDN) to capture local geometric patterns associated with dynamically stable crystals. For each periodic neighboring atomic pair $(u,v)$, a pair representation is constructed from the initial atom embeddings before graph message passing,
\begin{equation}
\mathbf{e}_{uv}
=
\left[
\mathbf{x}_u,
\mathbf{x}_v,
|\mathbf{x}_u-\mathbf{x}_v|,
\mathbf{x}_u\odot\mathbf{x}_v
\right],
\end{equation}
where $\mathbf{x}_u$ denotes the initial embedding of atom $u$, and $\odot$ denotes element-wise multiplication. Using initial atom embeddings allows the geometry branch to directly learn local pairwise distance distributions conditioned on atomic species. The MDN predicts a mixture of $K$ Gaussian components conditioned on $\mathbf{e}_{uv}$,
\begin{equation}
\{\pi_k,\mu_k,\sigma_k\}_{k=1}^{K}
=
f_{\rm MDN}(\mathbf{e}_{uv}),
\end{equation}
where $\pi_k$, $\mu_k$, and $\sigma_k$ denote the mixture weight, mean, and standard deviation of the $k$-th Gaussian component, respectively. The likelihood of the observed pair distance $d_{uv}$ is computed as
\begin{equation}
p(d_{uv}\mid\mathbf{e}_{uv})
=
\sum_{k=1}^{K}
\pi_k
\,
\mathcal{N}
\!\left(
d_{uv};
\mu_k,
\sigma_k^2
\right).
\end{equation}
The graph-level geometry score is defined as the average log-likelihood over all periodic neighboring edges,
\begin{equation}
S_{\rm geom}
=
\frac{1}{|E|}
\sum_{(u,v)\in E}
\log
p(d_{uv}\mid\mathbf{e}_{uv}),
\end{equation}
where $E$ denotes the set of periodic neighbor edges in the crystal graph. Higher values of $S_{\rm geom}$ indicate that the local coordination geometry is more consistent with the geometric patterns learned from dynamically stable structures. To prevent strongly unstable structures from dominating geometry learning, the MDN likelihood is weighted according to the ground-truth minimum phonon frequency during training,
\begin{equation}
w_i
=
\sigma
\!\left(
\frac{\omega_i-\tau_{\rm stable}}{T}
\right),
\end{equation}
where $\tau_{\rm stable}$ denotes the stability threshold and $T$ is a temperature parameter controlling the transition. The MDN loss is defined as
\begin{equation}
L_{\rm mdn}
=
-\frac{
\sum_i
w_i
\sum_{(u,v)\in E_i}
\log
p(d_{uv}\mid\mathbf{e}_{uv})
}{
\sum_i
w_i|E_i|
},
\end{equation}
which encourages the geometry branch to primarily model local coordination patterns characteristic of dynamically stable structures. In addition, the graph-level geometry score is directly supervised using the threshold-aware ranking objective to improve its consistency with dynamical-stability ranking.

\subsection*{Unified scoring and ranking}
During training, the outputs of the three branches are combined into a unified ranking score,
\begin{equation}
S
=
S_{\rm reg}
+
\beta
S_{\rm thr}
+
\alpha
S_{\rm geom},
\end{equation}
where $S_{\rm reg}$, $S_{\rm thr}$, and $S_{\rm geom}$ denote the regression, threshold-classification, and geometry scores, respectively, while $\alpha$ and $\beta$ control the contributions of the geometry and classification branches. To directly optimize candidate prioritization, PhononScore is trained using a threshold-aware pairwise ranking objective. For a pair of structures $(i,j)$, the pairwise importance is defined by the number of stability thresholds separating the two structures,

\begin{equation}
w_{ij}
=
\sum_k
\max
(y_{i,k}-y_{j,k},0).
\end{equation}
The ranking loss is formulated as
\begin{equation}
L_{\rm rank}
=
\frac{
\sum_{i,j}
w_{ij}
\,
{\rm softplus}
\!\left(
-(S_i-S_j)
\right)
}{
\sum_{i,j}
w_{ij}
},
\end{equation}
which encourages structures satisfying more stringent stability criteria to consistently receive higher ranking scores, thereby directly optimizing stable-structure enrichment for high-throughput crystal screening. The overall training objective combines the regression, classification, ordinal-consistency, geometry-learning, geometry-ranking, and ranking objectives,
\begin{equation}
L
=
L_{\rm reg}
+
\lambda_{\rm cls}L_{\rm cls}
+
\lambda_{\rm ord}L_{\rm ord}
+
\lambda_{\rm mdn}L_{\rm mdn}
+
\lambda_{\rm geom}L_{\rm geom-rank}
+
\lambda_{\rm rank}L_{\rm rank},
\end{equation}
where $L_{\rm reg}$, $L_{\rm cls}$, $L_{\rm ord}$, $L_{\rm mdn}$, and $L_{\rm geom-rank}$ denote the regression, threshold-classification, ordinal-consistency, MDN likelihood, and geometry-ranking losses, respectively. Model training follows a two-stage strategy. In the first stage, PhononScore is pretrained on the large-scale MatterSim-labeled dataset consisting of generated structures and experimentally known materials. This stage enables the model to learn general structural patterns associated with dynamical stability across diverse crystal systems. In the second stage, the pretrained model is further fine-tuned using high-fidelity DFT-PBE phonon data, resulting in PhononScore-DFT. This stage calibrates the learned stability representations toward first-principles phonon calculations while retaining the statistical advantages of large-scale pretraining. 

During inference, the outputs of the three prediction branches are first standardized to remove scale differences,

\begin{equation}
z_i
=
\frac{S_i-\mu_i}{\sigma_i},
\end{equation}
where the mean $\mu_i$ and standard deviation $\sigma_i$ are computed within the candidate pool being evaluated. The final PhononScore used for candidate reranking is then computed as

\begin{equation}
S_{\rm PhononScore}
=
z_{\rm reg}
+
\beta
z_{\rm thr}
+
\alpha
z_{\rm geom},
\end{equation}
where the weighting coefficients $\alpha$ and $\beta$ are selected on the validation set and subsequently kept fixed for all benchmark evaluations.

To evaluate stable-structure enrichment, we primarily report the Top-$K$ stable rate and the enrichment factor. A structure is considered dynamically stable if its minimum phonon frequency satisfies $\omega_{\min}\ge\tau$, where $\tau=-0.1$~THz. The Top-$K$ stable rate is defined as
\begin{equation}
\mathrm{SR@}K
=
\frac{1}{K}
\sum_{i\in\mathrm{Top}K}
\mathbb{I}(\omega_i\ge\tau),
\end{equation}
where $\mathrm{Top}K$ denotes the $K$ highest-ranked candidates. Unless otherwise specified, we report $\mathrm{SR@100}$ throughout the main text. To quantify the improvement over the original candidate pool, we further introduce the enrichment factor,
\begin{equation}
\mathrm{EF@}K
=
\frac{\mathrm{SR@}K}
{\mathrm{SR}_{\rm pool}},
\end{equation}
where
\begin{equation}
\mathrm{SR}_{\rm pool}
=
\frac{1}{N}
\sum_{i=1}^{N}
\mathbb{I}(\omega_i\ge\tau)
\end{equation}
where $\mathrm{SR}_{\rm pool}$ is the stable fraction of the original candidate pool. $\mathrm{EF}>1$ indicates that reranking enriches dynamically stable structures relative to the input candidate pool. The enrichment factor, a standard metric in virtual screening and drug discovery~\cite{rtmscore,AlphaFold-3}, is introduced here as a quantitative metric for crystal screening to quantify the enrichment of dynamically stable structures after reranking.

\subsection*{DFT calculations}
DFT calculations were performed using the Vienna Ab initio Simulation Package (VASP)~\cite{vasp96prb,vasp96cms} with the PBE exchange--correlation functional within the generalized gradient approximation. All candidate structures were fully relaxed until the total energy and atomic forces converged to $10^{-8}$ eV per cell and $10^{-8}$ eV/\AA, respectively. After structural optimization, harmonic phonon spectra were calculated using the finite-displacement method as implemented in the PHONOPY package~\cite{phonopy-phono3py-JPSJ,phonopy-phono3py-JPCM}.

\section*{Data availability}

All datasets generated and analyzed during this study are publicly available through Zenodo:
\url{https://zenodo.org/records/21157982}. The repository includes the PhononScore benchmark datasets, associated metadata, and files required to reproduce the experiments reported in this work.

\section*{Code availability}
The source code, pretrained models, and evaluation scripts for PhononScore are publicly available at: \url{https://github.com/xqh19970407/PhononScore}. The repository contains the complete implementation of the PhononScore framework, training and inference pipelines, and scripts for reproducing the benchmark results presented in this paper. The PhononScore web platform is available at \url{http://phononbench.cn/phononscore/},
and the companion phonon calculation service is available at \url{http://phononbench.cn}.

\bibliographystyle{unsrt}
\bibliography{references}

@inproceedings{MACE-mp-0,
 author = {Batatia, Ilyes and Kovacs, David P and Simm, Gregor and Ortner, Christoph and Csanyi, Gabor},
 booktitle = {Advances in Neural Information Processing Systems},
 editor = {S. Koyejo and S. Mohamed and A. Agarwal and D. Belgrave and K. Cho and A. Oh},
 pages = {11423--11436},
 publisher = {Curran Associates, Inc.},
 title = {MACE: Higher Order Equivariant Message Passing Neural Networks for Fast and Accurate Force Fields},
 url = {https://proceedings.neurips.cc/paper_files/paper/2022/file/4a36c3c51af11ed9f34615b81edb5bbc-Paper-Conference.pdf},
 volume = {35},
 year = {2022}
}

@article{PhononBench-MP40,
	author={Li, Wen-Kao and Gao, Ze-Feng and Lu, Zhong-Yi},
	title={PhononBench-MP40: a spectrum-resolved benchmark dataset for phonon stability},
	journal={Chinese Physics B},
	url={http://iopscience.iop.org/article/10.1088/1674-1056/ae843f},
	year={2026}
}

@article{prb-bowen,
  title = {Superconductivity in atom-intercalated quaternary hydrides under ambient pressure},
  author = {Yao, Bo-Wen and Ouyang, Zhenfeng and Han, Xiao-Qi and Wu, Chang-Jiang and Guo, Peng-Jie and Gao, Ze-Feng and Lu, Zhong-Yi},
  journal = {Phys. Rev. B},
  volume = {113},
  issue = {9},
  pages = {094509},
  numpages = {8},
  year = {2026},
  month = {Mar},
  publisher = {American Physical Society},
  doi = {10.1103/rltl-vgzj},
  url = {https://link.aps.org/doi/10.1103/rltl-vgzj}
}

@misc{PhononBench,
      title={PhononBench:A Large-Scale Phonon-Based Benchmark for Dynamical Stability in Crystal Generation}, 
      author={Xiao-Qi Han and Peng-Jie Guo and Ze-Feng Gao and Zhong-Yi Lu},
      year={2025},
      eprint={2512.21227},
      archivePrefix={arXiv},
      primaryClass={cond-mat.mtrl-sci},
      url={https://arxiv.org/abs/2512.21227}, 
}

@misc{crysvcd,
      title={Enhancing Materials Discovery with Valence Constrained Design in Generative Modeling}, 
      author={Mouyang Cheng and Weiliang Luo and Hao Tang and Bowen Yu and Yongqiang Cheng and Weiwei Xie and Ju Li and Heather J. Kulik and Mingda Li},
      year={2025},
      eprint={2507.19799},
      archivePrefix={arXiv},
      primaryClass={cond-mat.mtrl-sci},
      url={https://arxiv.org/abs/2507.19799}, 
}

@article{m3gnet,
  author    = {Chen, Chi and Ong, Shyue Ping},
  title     = {A universal graph deep learning interatomic potential for the periodic table},
  journal   = {Nature Computational Science},
  year      = {2022},
  volume    = {2},
  number    = {11},
  pages     = {718--728},
  doi       = {10.1038/s43588-022-00349-3},
  url       = {https://doi.org/10.1038/s43588-022-00349-3}
}

@article{Deng2023CHGNet,
  author    = {Deng, Bowen and Zhong, Peichen and Jun, KyuJung and Riebesell, Janosh and Han, Kevin and Bartel, Christopher J. and Ceder, Gerbrand},
  title     = {CHGNet as a pretrained universal neural network potential for charge-informed atomistic modelling},
  journal   = {Nature Machine Intelligence},
  year      = {2023},
  volume    = {5},
  number    = {9},
  pages     = {1031--1041},
  doi       = {10.1038/s42256-023-00716-3},
  url       = {https://doi.org/10.1038/s42256-023-00716-3}
}

@article{pet-oam,
  author    = {Mazitov, Arslan and Bigi, Filippo and Kellner, Matthias and Pegolo, Paolo and Tisi, Davide and Fraux, Guillaume and Pozdnyakov, Sergey and Loche, Philip and Ceriotti, Michele},
  title     = {PET-MAD as a lightweight universal interatomic potential for advanced materials modeling},
  journal   = {Nature Communications},
  year      = {2025},
  volume    = {16},
  number    = {1},
  pages     = {10653},
  doi       = {10.1038/s41467-025-65662-7},
  url       = {https://doi.org/10.1038/s41467-025-65662-7}
}

@misc{eSEN-30M-OAM,
      title={Learning Smooth and Expressive Interatomic Potentials for Physical Property Prediction}, 
      author={Xiang Fu and Brandon M. Wood and Luis Barroso-Luque and Daniel S. Levine and Meng Gao and Misko Dzamba and C. Lawrence Zitnick},
      year={2025},
      eprint={2502.12147},
      archivePrefix={arXiv},
      primaryClass={physics.comp-ph},
      url={https://arxiv.org/abs/2502.12147}, 
}

@article{mattergen,
  author       = {Zeni, Claudio and Pinsler, Robert and Zügner, Daniel and Fowler, Andrew and Horton, Matthew and Fu, Xiang and Wang, Zilong and Shysheya, Aliaksandra and Crabbé, Jonathan and Ueda, Shoko and Sordillo, Roberto and Sun, Lixin and Smith, Jake and Nguyen, Bichlien and Schulz, Hannes and Lewis, Sarah and Huang, Chin-Wei and Lu, Ziheng and Zhou, Yichi and Yang, Han and Hao, Hongxia and Li, Jielan and Yang, Chunlei and Li, Wenjie and Tomioka, Ryota and Xie, Tian},
  title        = {A generative model for inorganic materials design},
  journal      = {Nature},
  year         = {2025},
  volume       = {639},
  number       = {8055},
  pages        = {624--632},
  doi          = {10.1038/s41586-025-08628-5},
  url          = {https://doi.org/10.1038/s41586-025-08628-5},
  issn         = {1476-4687}
}

@article{gnome,
  author    = {Merchant, Amil and Batzner, Simon and Schoenholz, Samuel S. and Aykol, Muratahan and Cheon, Gowoon and Cubuk, Ekin Dogus},
  title     = {Scaling deep learning for materials discovery},
  journal   = {Nature},
  year      = {2023},
  volume    = {624},
  number    = {7990},
  pages     = {80--85},
  doi       = {10.1038/s41586-023-06735-9},
  url       = {https://doi.org/10.1038/s41586-023-06735-9},
  issn      = {1476-4687},
  date      = {2023/12/01}
}

@article{invdesreview,
title = {{AI-Driven Inverse Design of Materials: Past, Present, and Future}},
journal = {Chinese Physics Letters},
volume = {42},
number = {2},
pages = {027403},
year = {2025},
issn = {},
doi = {10.1088/0256-307X/42/2/027403},	
url = {http://cpl.iphy.ac.cn/en/article/doi/10.1088/0256-307X/42/2/027403},
author = {Xiao-Qi Han and Xin-De Wang and Meng-Yuan Xu and Zhen Feng and Bo-Wen Yao and Peng-Jie Guo and Ze-Feng Gao and Zhong-Yi Lu}
}

@article{CrystalFlow,
  author    = {Xiaoshan Luo and Zhenyu Wang and Qingchang Wang and Xuechen Shao and Jian Lv and Lei Wang and Yanchao Wang and Yanming Ma},
  title     = {CrystalFlow: a flow-based generative model for crystalline materials},
  journal   = {Nature Communications},
  year      = {2025},
  volume    = {16},
  number    = {1},
  pages     = {9267},
  doi       = {10.1038/s41467-025-64364-4},
  url       = {https://doi.org/10.1038/s41467-025-64364-4}
}

@inproceedings{
diffcsp,
title={{Crystal Structure Prediction by Joint Equivariant Diffusion on Lattices and Fractional Coordinates}},
author={Rui Jiao and Wenbing Huang and Peijia Lin and Jiaqi Han and Pin Chen and Yutong Lu and Yang Liu},
booktitle={Workshop on ''Machine Learning for Materials'' ICLR 2023},
year={2023},
url={https://openreview.net/forum?id=VPByphdu24j}
}

@article{InvDesFlow-AL,
  author = {Xiao-Qi Han and Peng-Jie Guo and Ze-Feng Gao and Hao Sun and Zhong-Yi Lu},
  title = {InvDesFlow-AL: active learning-based workflow for inverse design of functional materials},
  journal = {npj Computational Materials},
  year = {2025},
  volume = {11},
  number = {1},
  pages = {364},
  doi = {10.1038/s41524-025-01830-z},
  url = {https://doi.org/10.1038/s41524-025-01830-z},
  issn = {2057-3960},
  date = {2025/11/24}
}

@misc{mattersim,
      title={MatterSim: A Deep Learning Atomistic Model Across Elements, Temperatures and Pressures}, 
      author={Han Yang and Chenxi Hu and Yichi Zhou and Xixian Liu and Yu Shi and Jielan Li and Guanzhi Li and Zekun Chen and Shuizhou Chen and Claudio Zeni and Matthew Horton and Robert Pinsler and Andrew Fowler and Daniel Zügner and Tian Xie and Jake Smith and Lixin Sun and Qian Wang and Lingyu Kong and Chang Liu and Hongxia Hao and Ziheng Lu},
      year={2024},
      eprint={2405.04967},
      archivePrefix={arXiv},
      primaryClass={cond-mat.mtrl-sci},
      url={https://arxiv.org/abs/2405.04967}, 
}

@misc{DPA4,
      title={DPA4: Pushing the Accuracy-Cost Frontier of Interatomic Potentials with EMFA SO(2) Convolution}, 
      author={Tiancheng Li and Wentao Li and Anyang Peng and Jianming Xue and Linfeng Zhang and Duo Zhang and Han Wang},
      year={2026},
      eprint={2606.02419},
      archivePrefix={arXiv},
      primaryClass={physics.chem-ph},
      url={https://arxiv.org/abs/2606.02419}, 
}

@article{rtmscore,
  author  = {Shen, Chao and Zhang, Xujun and Deng, Yafeng and Gao, Junbo and Wang, Dong and Xu, Lei and Pan, Peichen and Hou, Tingjun and Kang, Yu},
  title   = {Boosting Protein--Ligand Binding Pose Prediction and Virtual Screening Based on Residue--Atom Distance Likelihood Potential and Graph Transformer},
  journal = {Journal of Medicinal Chemistry},
  year    = {2022},
  volume   = {65},
  number   = {15},
  pages    = {10691--10706},
  doi      = {10.1021/acs.jmedchem.2c00991},
  url      = {https://doi.org/10.1021/acs.jmedchem.2c00991}
}

@article{AlphaFold-3,
  author  = {Abramson, Josh and Adler, Jonas and Dunger, Jack and Evans, Richard and Green, Tim and Pritzel, Alexander and Ronneberger, Olaf and Willmore, Lindsay and Ballard, Andrew J. and Bambrick, Joshua and Bodenstein, Sebastian and Evans, David A. and Ferris, Adam C. and Good, Isaac and Hamrick, James B. and Hassabis, Demis and Kohli, Pushmeet and Meyer, Clemens and O'Neill, Michael and Romera-Paredes, Bernardino and Sabbadini, Matteo and Simonyan, Karen and Vinyals, Oriol and Senior, Andrew W. and Kavukcuoglu, Koray and Jumper, John and Hassabis, Demis},
  title   = {Accurate Structure Prediction of Biomolecular Interactions with AlphaFold 3},
  journal = {Nature},
  year    = {2024},
  volume  = {630},
  number  = {8016},
  pages   = {493--500},
  doi     = {10.1038/s41586-024-07487-w},
  url     = {https://doi.org/10.1038/s41586-024-07487-w}
}

@article{Flex-jmc,
  author    = {Chao Shen and Xiaoqi Han and Heng Cai and Tong Chen and Yu Kang and Peichen Pan and Xiangyang Ji and Chang-Yu Hsieh and Yafeng Deng and Tingjun Hou},
  title     = {Improving the Reliability of Language Model-Predicted Structures as Docking Targets through Geometric Graph Learning},
  journal   = {Journal of Medicinal Chemistry},
  year      = {2025},
  volume     = {68},
  number     = {2},
  pages      = {1956--1969},
  doi        = {10.1021/acs.jmedchem.4c02740},
  url        = {https://doi.org/10.1021/acs.jmedchem.4c02740},
  publisher  = {American Chemical Society}
}

@article{Eberhardt2021Vina,
  author  = {Eberhardt, Jerome and Santos-Martins, Diogo and Tillack, Andreas F. and Forli, Stefano},
  title   = {AutoDock Vina 1.2.0: New Docking Methods, Expanded Force Field, and Python Bindings},
  journal = {Journal of Chemical Information and Modeling},
  year    = {2021},
  volume   = {61},
  number   = {8},
  pages    = {3891--3898},
  doi      = {10.1021/acs.jcim.1c00203},
  url      = {https://doi.org/10.1021/acs.jcim.1c00203}
}

@article{glide,
  author  = {Friesner, Richard A. and Banks, Jay L. and Murphy, Robert B. and Halgren, Thomas A. and Klicic, Jasna J. and Mainz, Daniel T. and Repasky, Matthew P. and Knoll, Eric H. and Shelley, Mee and Perry, Jason K. and Shaw, David E. and Francis, Perry and Shenkin, Peter S.},
  title   = {Glide: A New Approach for Rapid, Accurate Docking and Scoring. 1. Method and Assessment of Docking Accuracy},
  journal = {Journal of Medicinal Chemistry},
  year    = {2004},
  volume  = {47},
  number  = {7},
  pages   = {1739--1749},
  doi     = {10.1021/jm0306430}
}

@article{AutoDock,
author = {Trott, Oleg and Olson, Arthur J.},
title = {AutoDock Vina: Improving the speed and accuracy of docking with a new scoring function, efficient optimization, and multithreading},
journal = {Journal of Computational Chemistry},
volume = {31},
number = {2},
pages = {455-461},
doi = {https://doi.org/10.1002/jcc.21334},
url = {https://onlinelibrary.wiley.com/doi/abs/10.1002/jcc.21334},
eprint = {https://onlinelibrary.wiley.com/doi/pdf/10.1002/jcc.21334},
year = {2010}
}

@article{vasp96prb,
  title = {Efficient iterative schemes for ab initio total-energy calculations using a plane-wave basis set},
  author = {Kresse, G. and Furthm\"uller, J.},
  journal = {Phys. Rev. B},
  volume = {54},
  issue = {16},
  pages = {11169--11186},
  numpages = {0},
  year = {1996},
  month = {Oct},
  publisher = {American Physical Society},
  doi = {10.1103/PhysRevB.54.11169},
  url = {https://link.aps.org/doi/10.1103/PhysRevB.54.11169}
}

@article{vasp96cms,
title = {Efficiency of ab-initio total energy calculations for metals and semiconductors using a plane-wave basis set},
journal = {Computational Materials Science},
volume = {6},
number = {1},
pages = {15-50},
year = {1996},
issn = {0927-0256},
doi = {https://doi.org/10.1016/0927-0256(96)00008-0},
url = {https://www.sciencedirect.com/science/article/pii/0927025696000080},
author = {G. Kresse and J. Furthmüller},
}

@misc{DimeNet,
      title={Fast and Uncertainty-Aware Directional Message Passing for Non-Equilibrium Molecules}, 
      author={Johannes Gasteiger and Shankari Giri and Johannes T. Margraf and Stephan Günnemann},
      year={2022},
      eprint={2011.14115},
      archivePrefix={arXiv},
      primaryClass={cs.LG},
      url={https://arxiv.org/abs/2011.14115}, 
}

@article{CrysToGraph,
author = {Wang, Hongyi and Sun, Ji and Liang, Jinzhe and Zhai, Li and Tang, Zitian and Li, Zijian and Zhai, Wei and Wang, Xusheng and Gao, Weihao and Gong, Sheng},
title = {CrysToGraph: A Comprehensive Predictive Model for Crystal Material Properties and the Benchmark},
journal = {Battery Energy},
volume = {4},
number = {4},
pages = {e70004},
doi = {https://doi.org/10.1002/bte2.70004},
url = {https://onlinelibrary.wiley.com/doi/abs/10.1002/bte2.70004},
eprint = {https://onlinelibrary.wiley.com/doi/pdf/10.1002/bte2.70004},
year = {2025}
}

@article{DPA3,
  author  = {Zhang, Duo and Peng, Anyang and Cai, Chun and Li, Wentao and Zhou, Yuanchang and Zeng, Jinzhe and Guo, Mingyu and Zhang, Chengqian and Li, Bowen and Jiang, Hong and Zhu, Tong and Jia, Weile and Zhang, Linfeng and Wang, Han},
  title   = {A graph neural network for the era of large atomistic models},
  journal = {npj Computational Materials},
  year    = {2026},
  doi     = {10.1038/s41524-026-02146-2},
  url     = {https://doi.org/10.1038/s41524-026-02146-2}
}

@article{Choudhary2021,
  author  = {Choudhary, Kamal and DeCost, Brian},
  title   = {Atomistic Line Graph Neural Network for Improved Materials Property Predictions},
  journal = {npj Computational Materials},
  year    = {2021},
  volume  = {7},
  number  = {1},
  pages   = {185},
  doi     = {10.1038/s41524-021-00650-1},
  url     = {https://doi.org/10.1038/s41524-021-00650-1}
}

@misc{MDRPhononDB,
  author       = {{National Institute for Materials Science (NIMS)}},
  title        = {MDR Phonon Calculation Database},
  year         = {2024},
  howpublished = {\url{https://mdr.nims.go.jp/collections/8g84ms862?locale=en}},
  note         = {Accessed: 2024-11-04}
}

@article{mattersim-phonon,
  author = {Antoine Loew and Dewen Sun and Hai-Chen Wang and Silvana Botti and Miguel A. L. Marques},
  title = {Universal machine learning interatomic potentials are ready for phonons},
  journal = {npj Computational Materials},
  year = {2025},
  volume = {11},
  number = {1},
  pages = {178},
  doi = {10.1038/s41524-025-01650-1},
  url = {https://doi.org/10.1038/s41524-025-01650-1},
  issn = {2057-3960},
  date = {2025/06/12}
}

@article{MEGNet,
  title   = {Graph Networks as a Universal Machine Learning Framework for Molecules and Crystals},
  author  = {Chen, Chi and Ye, Weike and Zuo, Yunxing and Zheng, Chen and Ong, Shyue Ping},
  journal = {Chemistry of Materials},
  volume  = {31},
  number  = {9},
  pages   = {3564--3572},
  year    = {2019},
  publisher = {American Chemical Society},
  doi     = {10.1021/acs.chemmater.9b01294},
  url     = {https://doi.org/10.1021/acs.chemmater.9b01294}
}

@article{phonopy-phono3py-JPSJ,
  author  = {Togo, Atsushi},
  title   = {First-principles Phonon Calculations with Phonopy and Phono3py},
  journal = {J. Phys. Soc. Jpn.},
  volume  = {92},
  number  = {1},
  pages   = {012001},
  year    = {2023},
  doi     = {10.7566/JPSJ.92.012001}
}

@article{phonopy-phono3py-JPCM,
  author  = {Togo, Atsushi and Chaput, Laurent and Tadano, Terumasa and Tanaka, Isao},
  title   = {Implementation strategies in phonopy and phono3py},
  journal = {J. Phys. Condens. Matter},
  volume  = {35},
  number  = {35},
  pages   = {353001},
  year    = {2023},
  doi     = {10.1088/1361-648X/acd831}
}

@article{dpa2,
  author    = {Duo Zhang and Xinzijian Liu and Xiangyu Zhang and others},
  title     = {{DPA-2: a large atomic model as a multi-task learner}},
  journal   = {npj Computational Materials},
  year      = {2024},
  volume    = {10},
  number    = {1},
  pages     = {293},
  doi       = {10.1038/s41524-024-01493-2},
  url       = {https://doi.org/10.1038/s41524-024-01493-2}
}

@Article{InvDesFlow,
title = {{InvDesFlow: An AI-driven materials inverse design workflow to explore possible high-temperature superconductors}},
journal = {Chin. Phys. Lett.},
volume = {42},
number = {4},
pages = {047301},
year = {2025},
issn = {},
doi = {10.1088/0256-307X/42/4/047301},	
url = {http://cpl.iphy.ac.cn/en/article/doi/10.1088/0256-307X/42/4/047301},
author = {Xiao-Qi Han and Zhenfeng Ouyang and Peng-Jie Guo and Hao Sun and Ze-Feng Gao and Zhong-Yi Lu}
}

@article{liauh,
  title = {{High-temperature superconductivity in ${\mathrm{Li}}_{2}{\mathrm{AuH}}_{6}$ mediated by strong electron-phonon coupling under ambient pressure}},
  author = {Ouyang, Zhenfeng and Yao, Bo-Wen and Han, Xiao-Qi and Guo, Peng-Jie and Gao, Ze-Feng and Lu, Zhong-Yi},
  journal = {Phys. Rev. B},
  volume = {111},
  issue = {14},
  pages = {L140501},
  numpages = {6},
  year = {2025},
  month = {Apr},
  publisher = {American Physical Society},
  doi = {10.1103/PhysRevB.111.L140501},
  url = {https://link.aps.org/doi/10.1103/PhysRevB.111.L140501}
}

@article{alphafold3,
  author    = {Josh Abramson and Jonas Adler and Jack Dunger and Richard Evans},
  title     = {{Accurate structure prediction of biomolecular interactions with AlphaFold 3}},
  journal   = {Nature},
  year      = {2024},
  volume    = {630},
  number    = {8016},
  pages     = {493--500},
  doi       = {10.1038/s41586-024-07487-w},
  url       = {https://doi.org/10.1038/s41586-024-07487-w}
}

@article{crystalformer,
title = {Space group informed transformer for crystalline materials generation},
journal = {Science Bulletin},
volume = {70},
number = {21},
pages = {3522-3533},
year = {2025},
issn = {2095-9273},
doi = {https://doi.org/10.1016/j.scib.2025.09.035},
url = {https://www.sciencedirect.com/science/article/pii/S2095927325009752},
author = {Zhendong Cao and Xiaoshan Luo and Jian Lv and Lei Wang},
}

\vspace{36pt}
\noindent\textbf{Acknowledgement:}
The work is supported by Beijing Natural Science Foundation(No.Z250005), the National Natural Science Foundation of China (No.62476278, No.11934020), and the National Key R\&D Program of China (Grants No. 2024YFA1408601). Computational resources have been provided by the Physical Laboratory of High Performance Computing at Renmin University of China. 
\\

\noindent\textbf{Corresponding authors:} Correspondence and requests for materials should be addressed to Ze-Feng Gao (zfgao@ruc.edu.cn) and Zhong-Yi Lu (zlu@ruc.edu.cn). \\

\noindent\textbf{Competing interests:}
The authors declare no competing interests.\\

\noindent\textbf{Supplementary materials:}
The supplementary materials is attached.
% Extended Data Figure or Table
\clearpage

\appendix

\section*{Appendix Contents}
\begin{enumerate}
    \item \hyperref[app:ranking]{PhononScore Captures the Ranking of Dynamical Stability}

    \item \hyperref[app:threshold]{Robustness of PhononScore Across Stability Thresholds}

    \item \hyperref[app:weights]{Selection of the Standardized Score-Combination Weights}

    \item \hyperref[app:transfer]{Additional DFT-PBE Transfer Metrics}

    \item \hyperref[app:hardscreen]{Supplementary DFT Hard-Screening Analysis}

    \item \hyperref[app:cases]{Representative PhononScore-DFT Case Series}

    \item \hyperref[app:kio]{Additional K--I--O Phonon Diagnostics}

    \item \hyperref[app:split]{Dataset Splitting and Formula-Level Exclusion}

    \item \hyperref[app:training]{Training Objectives and Score Definitions}
\end{enumerate}

\section{PhononScore Captures the Ranking of Dynamical Stability}
\label{app:ranking}

\begin{figure*}[t!]
		\centering  
		\includegraphics[width=0.7\linewidth]{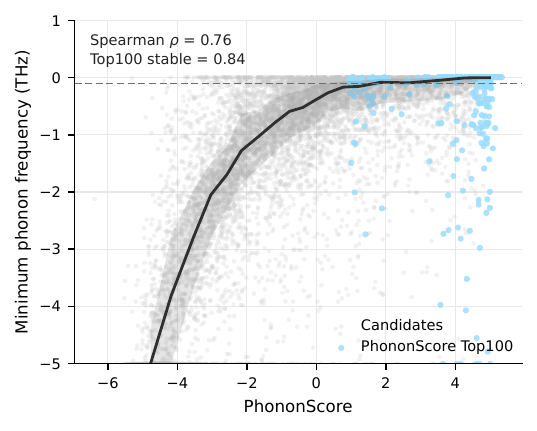}
		\caption{
\textbf{PhononScore captures a meaningful ordering of dynamical stability.}
Relationship between the final PhononScore and the true minimum phonon frequency ($\omega_{\min}$) for 8,000 generated crystal candidates. Gray points represent all candidates, while blue points denote the Top-100 candidates selected from each generative model according to PhononScore (800 structures in total). The black solid line shows the median $\omega_{\min}$ within score bins, and the shaded region indicates the interquartile range (25th--75th percentile). The horizontal dashed line marks the dynamical-stability threshold of $\omega_{\min}=-0.1$ THz. A clear monotonic trend is observed, with higher PhononScore values corresponding to larger minimum phonon frequencies. The selected high-scoring candidates are strongly enriched above the stability threshold, increasing the stability rate from 30.1\% in the original candidate pool to 83.8\% after reranking. The Spearman correlation coefficient between PhononScore and the true minimum phonon frequency is 0.758, demonstrating that PhononScore effectively captures a physically meaningful ranking signal associated with dynamical stability.
}
		\label{SM1-score-minfreq-scatter} 
\end{figure*}

Figure.~\ref{SM1-score-minfreq-scatter} illustrates the relationship between the final PhononScore and the true minimum phonon frequency $\omega_{\min}$ for all 8,000 generated crystal candidates collected from eight crystal generation models. Each gray point represents an individual candidate structure, with the horizontal axis corresponding to the predicted PhononScore and the vertical axis corresponding to the true minimum phonon frequency obtained from phonon calculations. To facilitate visualization, structures with extremely negative frequencies ($\omega_{\min}<-5$ THz) are clipped to $-5$ THz in the plot, while all statistical analyses are performed using the original, unclipped values.

A clear monotonic relationship is observed between the predicted score and the true minimum phonon frequency. The black solid curve shows the median $\omega_{\min}$ within score bins, while the shaded region denotes the interquartile range (25th--75th percentile). As the PhononScore increases, both the median minimum phonon frequency and the overall frequency distribution shift toward larger values. This trend indicates that structures assigned higher scores tend to exhibit fewer imaginary phonon modes and therefore possess stronger dynamical stability. The horizontal dashed line marks the dynamical-stability threshold of $\omega_{\min}=-0.1$ THz. Structures above this threshold are regarded as dynamically stable or near-stable, whereas structures below the threshold contain significant imaginary phonon modes and are considered dynamically unstable. Notably, candidates selected by PhononScore reranking are strongly enriched above this stability boundary. The blue points correspond to the Top-100 candidates selected from each generative model according to PhononScore (800 structures in total). Compared with the full candidate pool, these high-scoring structures are concentrated predominantly in the dynamically stable region, demonstrating that the score effectively prioritizes stable candidates for subsequent validation.

Quantitatively, the Spearman rank correlation coefficient between PhononScore and the true minimum phonon frequency reaches 0.758 across all 8,000 generated structures. This relatively strong rank correlation suggests that the score captures meaningful information related to phonon stability and preserves the relative ordering of candidates according to their likelihood of being dynamically stable. Importantly, the objective of PhononScore is not to predict the exact value of the minimum phonon frequency for every structure. Instead, it is designed as a ranking function that preferentially assigns higher scores to structures with higher stability. Consistent with this objective, the dynamical-stability rate increases substantially from 30.1\% in the original candidate pool to 83.8\% among the Top-100 candidates selected from each source model. This figure demonstrates that PhononScore learns a physically meaningful stability-related ranking signal rather than merely distinguishing structures generated by different models. The strong enrichment of dynamically stable structures among high-scoring candidates highlights its effectiveness as a reranking function for crystal generation pipelines and supports its use for efficiently prioritizing structures for expensive first-principles phonon calculations.

\section{Robustness of PhononScore Across Stability Thresholds}
\label{app:threshold}
To further evaluate the robustness of PhononScore, we repeated the reranking analysis under multiple dynamical-stability thresholds ranging from the nearly strict stability criterion ($\omega_{\min}>-0.001$ THz) to the much looser criterion ($\omega_{\min}>-1$ THz) (Table~\ref{tab:ablation_thresholds}). Across all crystal generators, PhononScore consistently outperformed direct ALIGNN-based ranking under the practically relevant thresholds of $-0.001$, $-0.01$, and $-0.1$ THz. The improvement was particularly pronounced near the stability boundary. For example, under the strictest threshold ($\omega_{\min}>-0.001$ THz), PhononScore improved the Top-100 stability rate from 0.53 to 0.92 for DiffCSP, from 0.46 to 0.82 for MatterGen, and from 0.49 to 0.81 for CrystalLLM+sg, corresponding to absolute gains of 0.39, 0.36, and 0.32, respectively.

Interestingly, the performance gap between ALIGNN and PhononScore systematically decreased as the stability threshold became more relaxed. At $\omega_{\min}>-1$ THz, both methods achieved similarly high selection accuracy, and several generators even exhibited marginal advantages for direct ALIGNN prediction. This behavior is expected because the loose threshold mainly distinguishes strongly unstable structures from the rest, a task that can already be captured reasonably well by directly predicting the minimum phonon frequency. In contrast, identifying structures close to the dynamical-stability boundary requires a more refined assessment of phonon-related characteristics, where the multi-task phonon representations learned by PhononScore become particularly beneficial.

These results demonstrate that the advantage of PhononScore does not arise merely from learning coarse stability trends. Instead, PhononScore is especially effective at ranking candidates near the physically meaningful stability boundary, which is precisely the regime most relevant for practical crystal discovery and high-throughput candidate screening.

\begin{table*}[t]
\centering
\small
\caption{
\textbf{Threshold-resolved ablation study of PhononScore.}
Values represent the true dynamical stability rate of the Top-100 structures selected from 1,000 held-out candidates for each generator under different stability thresholds. Dynamical stability is defined as $\omega_{\min}$ exceeding the specified threshold.
}
\label{tab:ablation_thresholds}
\begin{tabular}{llccc}
\toprule
Generator Source &
Threshold (THz) &
PhononBench (\%) &
ALIGNN ($\omega_{\min}$) (\%) &
PhononScore (\%) \\
\midrule

CrystalFlow & -0.001 & 16.8 & 48.0 & \textbf{88.0} \\
            & -0.01  & 17.6 & 48.0 & \textbf{88.0} \\
            & -0.1   & 40.1 & 74.0 & \textbf{92.0} \\
            & -1     & 82.6 & 97.0 & \textbf{98.0} \\
\midrule

CrystalFormer & -0.001 & 11.6 & 57.0 & \textbf{65.0} \\
              & -0.01  & 12.0 & 58.0 & \textbf{66.0} \\
              & -0.1   & 15.5 & 63.0 & \textbf{69.0} \\
              & -1     & 30.6 & \textbf{82.0} & 80.0 \\
\midrule

CrystalLLM-small & -0.001 & 3.0 & 20.0 & \textbf{26.0} \\
                 & -0.01  & 3.7 & 23.0 & \textbf{30.0} \\
                 & -0.1   & 14.4 & 50.0 & \textbf{60.0} \\
                 & -1     & 58.6 & \textbf{98.0} & 92.0 \\
\midrule

CrystalLLM-large & -0.001 & 18.4 & 65.0 & \textbf{88.0} \\
                 & -0.01  & 18.8 & 67.0 & \textbf{89.0} \\
                 & -0.1   & 23.4 & 73.0 & \textbf{90.0} \\
                 & -1     & 47.5 & 91.0 & \textbf{98.0} \\
\midrule

CrystalLLM+sg & -0.001 & 19.6 & 49.0 & \textbf{81.0} \\
              & -0.01  & 20.3 & 50.0 & \textbf{82.0} \\
              & -0.1   & 31.1 & 73.0 & \textbf{88.0} \\
              & -1     & 59.3 & 92.0 & \textbf{96.0} \\
\midrule

DiffCSP & -0.001 & 27.2 & 53.0 & \textbf{92.0} \\
         & -0.01  & 27.9 & 55.0 & \textbf{94.0} \\
         & -0.1   & 43.9 & 72.0 & \textbf{96.0} \\
         & -1     & 80.3 & 96.0 & \textbf{100.0} \\
\midrule

LLaMA2-70B & -0.001 & 21.7 & 61.0 & \textbf{82.0} \\
           & -0.01  & 22.5 & 64.0 & \textbf{85.0} \\
           & -0.1   & 32.5 & 77.0 & \textbf{85.0} \\
           & -1     & 57.8 & \textbf{96.0} & 94.0 \\
\midrule

InvDesFlow-AL & -0.001 & 27.2 & 53.0 & \textbf{92.0} \\
         & -0.01  & 27.9 & 55.0 & \textbf{94.0} \\
         & -0.1   & 43.9 & 72.0 & \textbf{96.0} \\
         & -1     & 80.3 & 96.0 & \textbf{100.0} \\
\midrule

MatterGen & -0.001 & 24.6 & 46.0 & \textbf{82.0} \\
          & -0.01  & 25.6 & 48.0 & \textbf{82.0} \\
          & -0.1   & 45.1 & 73.0 & \textbf{90.0} \\
          & -1     & 79.2 & 94.0 & \textbf{96.0} \\
\bottomrule
\end{tabular}
\end{table*}

\section{Selection of the Standardized Score-Combination Weights}
\label{app:weights}
In the main text, the evaluation-time PhononScore is defined as a standardized linear combination of three model components:
\begin{equation}
S_{\mathrm{eval}}
= z(S_{\mathrm{reg}})
+ \beta z(S_{\mathrm{thr}})
+ \alpha z(S_{\mathrm{geom}}),
\end{equation}
where $S_{\mathrm{reg}}$ is the regression-head output, $S_{\mathrm{thr}}$ is the multi-threshold stability score, and $S_{\mathrm{geom}}$ is the pair-geometry likelihood score. The standardization operator $z(\cdot)$ is applied within the candidate pool being evaluated. This normalization removes differences in numerical scale among the three components and allows the final score to focus on relative ranking within a screening pool.

\begin{figure*}[t!]
	\centering
	\includegraphics[width=0.68\linewidth]{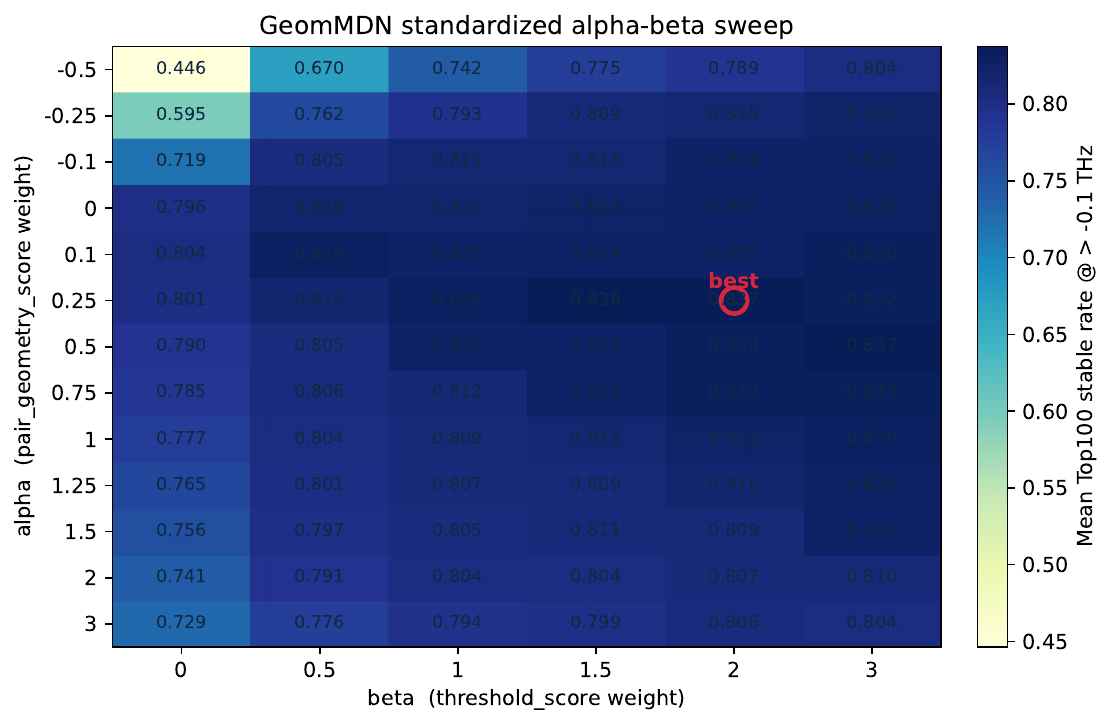}
	\caption{
	\textbf{Selection of the standardized score-combination weights.}
	Heatmap of the mean Top-100 dynamical-stability rate across the eight MP20 generation sources as a function of the pair-geometry weight $\alpha$ and multi-threshold stability weight $\beta$. The evaluation score is computed as $S_{\mathrm{eval}}=z(S_{\mathrm{reg}})+\beta z(S_{\mathrm{thr}})+\alpha z(S_{\mathrm{geom}})$. The selected setting, $\alpha=0.25$ and $\beta=2.0$, lies in a broad high-performing region and is fixed for all subsequent MP20 and DFT-PBE evaluations.
	}
	\label{SM2-alpha-beta-heatmap}
\end{figure*}

To select the fixed combination weights used throughout the paper, we performed a two-dimensional sweep over the geometry weight $\alpha$ and threshold weight $\beta$ on the MP20 generated-candidate validation benchmark. The objective was the mean Top-100 stability rate across the eight generation sources under the stability criterion $\omega_{\min}>-0.1$ THz. As shown in Fig.~\ref{SM2-alpha-beta-heatmap}, the stable-rate landscape contains a broad high-performing region, indicating that the final performance is not due to a single finely tuned point. We selected $\alpha=0.25$ and $\beta=2.0$, which achieved a mean Top-100 stability rate of 0.8375 on the MP20 generated-candidate benchmark. These weights were then fixed and directly reused for the DFT-PBE evaluations without further tuning on the DFT test set.

\section{Additional DFT-PBE Transfer Metrics}
\label{app:transfer}

\begin{figure*}[t!]
	\centering
	\includegraphics[width=0.68\linewidth]{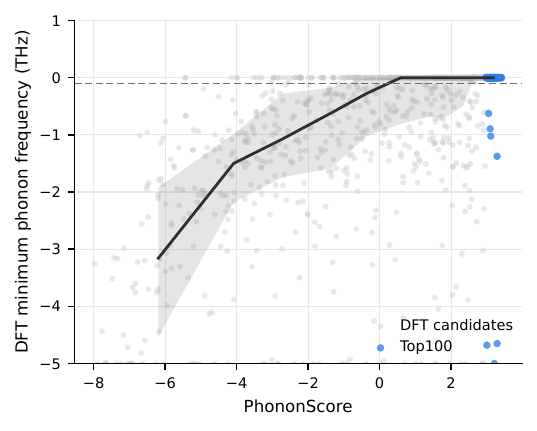}
	\caption{
	\textbf{Relationship between PhononScore-DFT and DFT-PBE minimum phonon frequency.}
	PhononScore-DFT shows a clear positive association with the true DFT-PBE minimum phonon frequency on the balanced 1,000-structure DFT test set, indicating that high-score candidates are enriched in dynamically stable or near-stable structures.
	}
	\label{SM3-dft-score-minfreq-scatter}
\end{figure*}

The main text demonstrates that PhononScore transfers from MatterSim-derived labels to high-fidelity DFT-PBE phonon labels and that DFT post-training further improves the ranking of DFT-stable structures. Here we provide additional diagnostic plots supporting this conclusion. Fig.~\ref{SM3-dft-score-minfreq-scatter} shows the relationship between the PhononScore-DFT score and the DFT-PBE minimum phonon frequency across the balanced 1,000-structure DFT test set. Although PhononScore-DFT is optimized as a ranking score rather than a calibrated frequency regressor, the distribution exhibits a clear positive trend: high-score structures are enriched in the region with larger minimum phonon frequencies and are therefore more likely to satisfy the dynamical-stability criterion.

\begin{figure*}[t!]
	\centering
	\includegraphics[width=0.5\linewidth]{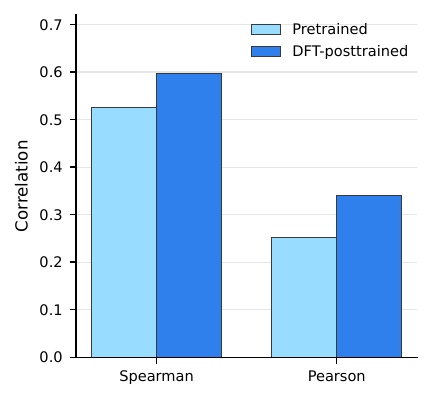}
	\caption{
	\textbf{Rank-correlation metrics for DFT-PBE transfer.}
	Comparison of rank-correlation metrics for the MatterSim-pretrained PhononScore and the DFT-posttrained PhononScore-DFT. DFT post-training improves the alignment between the scoring function and high-fidelity DFT-PBE phonon labels.
	}
	\label{SM3-dft-rank-correlation}
\end{figure*}

Fig.~\ref{SM3-dft-rank-correlation} compares rank-correlation metrics before and after DFT-PBE post-training. The pretrained PhononScore already provides a useful transfer signal, while PhononScore-DFT further improves the agreement with DFT phonon labels. This result supports the two-stage training strategy: large-scale MatterSim-derived phonon labels provide broad structural stability priors, and the smaller but higher-fidelity DFT-PBE dataset calibrates the score toward DFT-level stability ordering.

\section{Supplementary DFT Hard-Screening Analysis}
\label{app:hardscreen}
In practical materials discovery, dynamically stable candidates are often rare within a large generated pool. To mimic this imbalanced screening scenario, the main text evaluates PhononScore-DFT under hard-screening settings constructed by repeated resampling from the balanced DFT-PBE test set. Each hard pool contains 200 structures with a controlled stable fraction: 5\% (Extreme), 10\% (Strict), or 20\% (Moderate). The model is evaluated by reranking each pool and measuring the stable fraction within the top-ranked candidates.

\begin{figure*}[t!]
	\centering
	\includegraphics[width=0.5\linewidth]{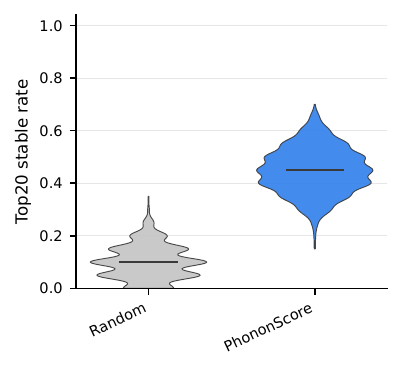}
	\caption{
	\textbf{Repeat-level distribution of hard-screening performance.}
	Distribution of the Top-20 stable rate across 1000 resampled hard-screening pools with 5\%, 10\%, and 20\% DFT-stable structures. PhononScore-DFT robustly enriches DFT-stable structures in the top-ranked region even when stable candidates are rare.
	}
	\label{SM4-dft-hard-repeat}
\end{figure*}

The Top-$K$ stable-rate curves for these hard-screening settings are shown in the main text. As an additional robustness check, Fig.~\ref{SM4-dft-hard-repeat} shows the distribution of Top-20 stable rates over 1000 repeated resampling trials, demonstrating that the enrichment is robust across random pool compositions rather than being driven by a small number of favorable draws. This supplementary analysis supports the use of PhononScore-DFT as a practical prescreening tool when DFT-stable structures are scarce.

\section{Representative PhononScore-DFT Case Series}
\label{app:cases}
To test whether PhononScore-DFT can recover nontrivial stability trends beyond simple chemical composition effects, we mined representative case series from the DFT post-training set. These series include polymorphs with the same composition, compounds with the same element set but different stoichiometries, and chemically related materials connected by anonymous-formula or element-family substitution. Table~\ref{tab:phononscore_dft_case_series} summarizes representative examples. In each series, the ordering of PhononScore-DFT is broadly consistent with the ordering of the DFT-PBE minimum phonon frequency, suggesting that the model captures stability variations associated with structural arrangement, stoichiometry, and chemically related substitutions.

\begin{table*}[t]
\centering
\caption{
Representative PhononScore-DFT case series from Materials Project. Materials within each series are ordered by DFT-PBE minimum phonon frequency. Case definitions: (I) same composition with different space groups/structures (RbPO$_3$ polymorphs); (II) same element set with different stoichiometries (Sr--Mg--H hydrides); (III) same element set with different stoichiometries (Rb--In--S sulfides); (IV) same space group and anonymous formula with element-family substitution (ABC$_2$ chalcogenides, SG 15); and (V) same anonymous formula with element-family substitution (alkali oxy-pnictogenates).
}
\label{tab:phononscore_dft_case_series}
\begin{tabular}{lllrrr}
\toprule
Case & MP ID & Formula & SG & $\omega_{\min}^{\mathrm{DFT}}$ (THz) & PhononScore-DFT \\
\midrule

\multirow{3}{*}{\textbf{Case I}}
& mp-548658 & RbPO$_3$ & 63 & -5.141 & -10.545 \\
& mp-9138   & RbPO$_3$ & 62 & -1.507 & -3.029 \\
& mp-4135   & RbPO$_3$ & 14 & -0.000 & 0.218 \\
\cmidrule(lr){1-6}

\multirow{3}{*}{\textbf{Case II}}
& mp-644225 & Sr$_2$MgH$_6$ & 164 & -7.004 & -10.938 \\
& mp-707440 & Sr$_2$Mg$_3$H$_{10}$ & 12 & -4.692 & -7.870 \\
& mp-643009 & SrMgH$_4$ & 36 & -0.000 & 1.652 \\
\cmidrule(lr){1-6}

\multirow{4}{*}{\textbf{Case III}}
& mp-601861 & RbInS$_2$ & 15 & -24.546 & -6.171 \\
& mp-20938  & RbIn$_5$S$_8$ & 12 & -0.539 & -1.331 \\
& mp-22303  & Rb$_3$InS$_3$ & 12 & -0.000 & 1.512 \\
& mp-542654 & RbIn$_3$S$_5$ & 10 & -0.000 & 0.115 \\
\cmidrule(lr){1-6}

\multirow{5}{*}{\textbf{Case IV}}
& mp-601861 & RbInS$_2$ & 15 & -24.546 & -6.171 \\
& mp-17650  & KGaS$_2$ & 15 & -0.000 & -4.688 \\
& mp-559459 & CsInS$_2$ & 15 & -0.000 & -3.556 \\
& mp-561407 & RbGaS$_2$ & 15 & -0.000 & 0.158 \\
& mp-5038   & CsGaS$_2$ & 15 & -0.000 & 2.124 \\
\cmidrule(lr){1-6}

\multirow{5}{*}{\textbf{Case V}}
& mp-548658 & RbPO$_3$ & 63 & -5.141 & -10.545 \\
& mp-9138   & RbPO$_3$ & 62 & -1.507 & -3.029 \\
& mp-557189 & LiPO$_3$ & 13 & -0.000 & 1.194 \\
& mp-4531   & NaNO$_3$ & 167 & -0.000 & 1.715 \\
& mp-770932 & LiSbO$_3$ & 12 & 0.000 & 2.172 \\

\bottomrule
\end{tabular}
\end{table*}

\section{Additional K--I--O Phonon Diagnostics}
\label{app:kio}
The K--I--O family discussed in the main text provides a compact example in which PhononScore-DFT distinguishes materials that share similar chemistry but differ in dynamical-stability behavior. The full phonon-band and projected density-of-states analyses show that KIO$_4$ is dynamically stable within the computed spectrum, whereas KIO$_3$ and K$_3$IO$_5$ exhibit imaginary modes. Here we provide two additional diagnostics that connect the score-level ranking to the underlying phonon instabilities.

\begin{figure*}[t!]
	\centering
	\includegraphics[width=0.68\linewidth]{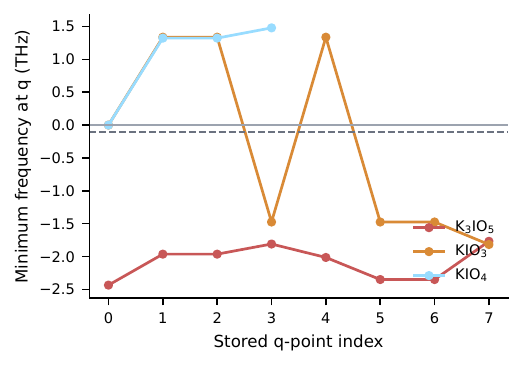}
	\caption{
	\textbf{Minimum phonon frequency along sampled reciprocal-space points for K--I--O structures.}
	The imaginary modes in KIO$_3$ and K$_3$IO$_5$ are localized in specific regions of the high-symmetry paths, indicating collective soft modes rather than a uniform shift of the entire phonon spectrum.
	}
	\label{SM5-kio-qpoint-minima}
\end{figure*}

Fig.~\ref{SM5-kio-qpoint-minima} summarizes the minimum phonon frequency resolved along the sampled reciprocal-space points. For KIO$_3$, the most unstable modes are concentrated around the selected high-symmetry path segment containing the $H$--$L$--$\Gamma$--$S$ region, indicating that the instability is not a uniform shift of all modes but a soft collective distortion localized in a specific part of the Brillouin zone. For K$_3$IO$_5$, the stronger imaginary branch appears close to the low-frequency part of the $\Gamma$--$X$--$M$--$\Gamma$ path, consistent with a more pronounced lattice-level soft mode. In both cases, the instability is therefore better interpreted as a collective dynamical effect rather than a simple single-variable chemical descriptor.

\begin{figure*}[t!]
	\centering
	\includegraphics[width=0.95\linewidth]{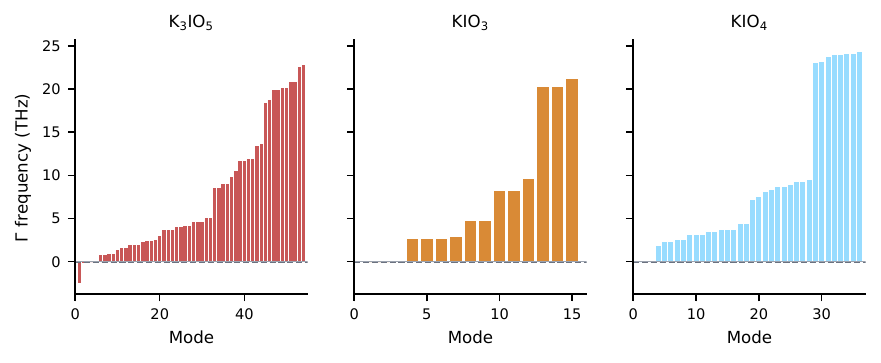}
	\caption{
	\textbf{Atom-resolved participation of low-frequency $\Gamma$-point modes in the K--I--O case study.}
	The low-frequency mode participation indicates that the imaginary branches in KIO$_3$ and K$_3$IO$_5$ involve collective lattice softening of the iodine--oxygen framework rather than a purely composition-level effect.
	}
	\label{SM5-kio-gamma-modes}
\end{figure*}

Fig.~\ref{SM5-kio-gamma-modes} compares the atom-resolved participation of the lowest $\Gamma$-point modes. Combined with the projected density of states, the imaginary branches in KIO$_3$ and K$_3$IO$_5$ contain substantial oxygen-related contributions, with iodine-centered polyhedral units participating through collective I--O framework distortions. This observation is consistent with the chemical intuition that the stability of iodine oxides is controlled by the coupling between local coordination polyhedra and long-wavelength lattice dynamics. PhononScore-DFT does not explicitly solve the dynamical matrix during inference, but its score ordering is consistent with these phonon-level diagnostics, suggesting that the learned structural representation captures stability signatures associated with local coordination and collective softness.

\section{Dataset Splitting and Formula-Level Exclusion}
\label{app:split}
To reduce information leakage between training and evaluation, all benchmark test structures were fixed using the random seed \texttt{seed20260605}, and the reduced formulae appearing in the test sets were excluded from the corresponding training pools. This formula-level exclusion is stricter than a random structure-level split because it prevents the model from seeing the same reduced chemical formula during training and testing, even if the specific crystal structure differs.

The full multi-fidelity phonon collection contains 157,463 structures, including 100,606 AI-generated structures labeled by MatterSim-driven phonon calculations, 46,899 MP40 structures labeled by the same MatterSim/Phonopy workflow, and 9,958 structures with DFT-PBE phonon labels. After formula exclusion, the pretraining pool contains 133,389 structures, comprising 88,480 generated structures and 44,909 MP40 structures. The DFT post-training pool contains 8,221 structures. The final held-out benchmark contains three test subsets: 8,000 generated structures from eight MP20 generation sources, 1,000 balanced DFT-PBE structures. The test structures cover 9,818 unique reduced formulae that are excluded from training.

\begin{table*}[t]
\small
\centering
\caption{
\textbf{Summary of the datasets and formula-level split used for PhononScore.}
Reduced formulae appearing in the held-out test sets were excluded from all training datasets.
}
\label{tab:dataset-split-summary}
\begin{tabular}{llr}
\toprule
Category & Dataset & Structures \\
\midrule
Full dataset & Generated & 100,606 \\
             & MP40 & 46,899 \\
             & DFT-PBE & 9,958 \\
\midrule
Training & Pretraining & 133,389 \\
         & Generated & 88,480 \\
         & MP40 & 44,909 \\
         & DFT fine-tuning & 8,221 \\
\midrule
Test & MP20 benchmark & 9,000 \\
     & DFT-PBE benchmark & 1,000 \\
     % & Alex20 benchmark & 1,000 \\
     % & Unique formulae & 9,818 \\
\bottomrule
\end{tabular}
\end{table*}

\section{Training Objectives and Score Definitions}
\label{app:training}
PhononScore is trained as a scoring function rather than as a pure phonon-frequency regressor. The model uses the ALIGNN periodic atom graph and line graph encoder to obtain structure-level and pair-level representations, and jointly optimizes regression, multi-threshold classification, pair-geometry likelihood, and ranking-oriented objectives.

For the regression head, the target is the minimum phonon frequency clipped to a finite range,
\begin{equation}
\tilde{\omega}_{i}
= \mathrm{clip}(\omega^{\min}_{i}, -5, 0),
\end{equation}
and the regression loss is
\begin{equation}
\mathcal{L}_{\mathrm{reg}}
= \frac{1}{N}\sum_{i=1}^{N}
\mathrm{SmoothL1}(\hat{\omega}_{i}, \tilde{\omega}_{i}).
\end{equation}
Here $\omega^{\min}_{i}$ is the true minimum phonon frequency and $\hat{\omega}_{i}$ is the model-predicted regression score.

For the multi-threshold stability head, each structure is assigned binary labels under several stability thresholds,
\begin{equation}
y_{i,k}=\mathbb{I}\left[\omega^{\min}_{i}> \tau_k\right],
\qquad
\tau_k \in \{-0.001,-0.01,-0.1,-1.0\}\ \mathrm{THz}.
\end{equation}
The classification loss is the class-balanced binary cross entropy used in the implementation,
\begin{equation}
\mathcal{L}_{\mathrm{cls}}
= -\frac{1}{NK}\sum_{i=1}^{N}\sum_{k=1}^{K}
\eta_{i,k}
\left[
y_{i,k}\log p_{i,k}
+(1-y_{i,k})\log(1-p_{i,k})
\right],
\end{equation}
where $p_{i,k}$ is the predicted stability probability at threshold $\tau_k$. The class-balance weight is $\eta_{i,k}=w_k^+$ for positive samples and $\eta_{i,k}=1$ for negative samples, with $w_k^+=(N_k^-+\epsilon)/(N_k^+ + \epsilon)$ computed within the mini-batch for each threshold. The threshold score used for ranking is the mean probability across thresholds,
\begin{equation}
S_{\mathrm{thr},i}
= \frac{1}{K}\sum_{k=1}^{K}p_{i,k}.
\end{equation}

The pair-geometry branch is designed to learn local geometric patterns that are statistically associated with stable structures. For an edge $(u,v)$ with pair representation $\bm{\phi}_{uv}$ and distance $d_{uv}$, the model predicts a mixture density,
\begin{equation}
p(d_{uv}\mid \bm{\phi}_{uv})
= \sum_{m=1}^{M}\pi_{uv,m}
\mathcal{N}(d_{uv};\mu_{uv,m},\sigma_{uv,m}^{2}).
\end{equation}
The pair-geometry score is the mean log-likelihood over local pairs,
\begin{equation}
S_{\mathrm{geom},i}
= \frac{1}{|\mathcal{E}_{i}|}
\sum_{(u,v)\in \mathcal{E}_{i}}
\log p(d_{uv}\mid \bm{\phi}_{uv}).
\end{equation}
During training, this likelihood is weighted toward dynamically stable structures using
\begin{equation}
w^{\mathrm{geom}}_i
= \sigma\left(\frac{\omega^{\min}_{i}-\tau_{\mathrm{stable}}}{T}\right),
\qquad
\tau_{\mathrm{stable}}=-0.1\ \mathrm{THz},\quad T=0.2,
\end{equation}
and the stable-weighted MDN loss is computed over periodic edges as
\begin{equation}
\mathcal{L}_{\mathrm{mdn}}
=
-\frac{
\sum_{i} w^{\mathrm{geom}}_i
\sum_{(u,v)\in \mathcal{E}_{i}}
\log p(d_{uv}\mid \bm{\phi}_{uv})
}{
\sum_i w^{\mathrm{geom}}_i |\mathcal{E}_{i}|
}.
\end{equation}
This normalization matches the edge-level implementation in which each edge inherits the stability weight of its parent graph. It makes the geometry branch emphasize local coordination patterns characteristic of stable candidates.

The overall training loss is
\begin{equation}
\mathcal{L}
= \mathcal{L}_{\mathrm{reg}}
+ \lambda_{\mathrm{cls}}\mathcal{L}_{\mathrm{cls}}
+ \lambda_{\mathrm{rank}}\mathcal{L}_{\mathrm{rank}}
+ \lambda_{\mathrm{ord}}\mathcal{L}_{\mathrm{ord}}
+ \lambda_{\mathrm{mdn}}\mathcal{L}_{\mathrm{mdn}}
+ \lambda_{\mathrm{geom\_rank}}\mathcal{L}_{\mathrm{geom\_rank}},
\end{equation}
where the default weights are
\begin{equation}
\lambda_{\mathrm{cls}}=0.3,\quad
\lambda_{\mathrm{rank}}=0.3,\quad
\lambda_{\mathrm{ord}}=0.02,\quad
\lambda_{\mathrm{mdn}}=0.05,\quad
\lambda_{\mathrm{geom\_rank}}=0.05.
\end{equation}

The internal training-time score is
\begin{equation}
S_{\mathrm{train},i}
= \hat{\omega}_{i}
+0.6 S_{\mathrm{thr},i}
+0.1 S_{\mathrm{geom},i}.
\end{equation}
For candidate-pool reranking at evaluation time, we use the standardized score
\begin{equation}
S_{\mathrm{eval},i}
= z(\hat{\omega}_{i})
+2.0 z(S_{\mathrm{thr},i})
+0.25 z(S_{\mathrm{geom},i}),
\end{equation}
where $z(\cdot)$ denotes z-score normalization within the candidate pool. For a single isolated CIF, $S_{\mathrm{eval}}$ is not uniquely defined because the normalization requires a pool; in that case, the individual score components should be inspected directly.

The primary screening metric is the Top-$K$ stable rate. Given a candidate pool $\mathcal{C}=\{x_i\}_{i=1}^{N}$, score $S_i$, and stability threshold $\tau$, the true stability label is
\begin{equation}
y_i(\tau)=\mathbb{I}\left[\omega^{\min}_{i}\ge \tau\right].
\end{equation}
After sorting candidates by decreasing $S_i$, the Top-$K$ stable rate is
\begin{equation}
\mathrm{SR@}K
= \frac{1}{K}\sum_{i\in \mathrm{Top}K(\mathcal{C})}
y_i(\tau).
\end{equation}
The pool-level stable rate is
\begin{equation}
\mathrm{SR}_{\mathrm{pool}}
= \frac{1}{N}\sum_{i=1}^{N}y_i(\tau),
\end{equation}
and the enrichment factor is
\begin{equation}
\mathrm{EF@}K
= \frac{\mathrm{SR@}K}{\mathrm{SR}_{\mathrm{pool}}}.
\end{equation}
Unless otherwise stated, the main screening threshold is $\tau=-0.1$ THz. A value of $\mathrm{EF@}K>1$ indicates that PhononScore enriches dynamically stable structures within the top-ranked candidates relative to the original pool.

\end{document}